\documentclass[sn-nature]{sn-jnl}

\usepackage{graphicx}%
\usepackage{multirow}%
\usepackage{amsmath,amssymb,amsfonts}%
\usepackage{amsthm}%
\usepackage{mathrsfs}%
\usepackage[title]{appendix}%
\usepackage{xcolor}%
\usepackage{textcomp}%
\usepackage{manyfoot}%
\usepackage{booktabs}%
\usepackage{algorithm}%
\usepackage{algorithmicx}%
\usepackage{algpseudocode}%
\usepackage{listings}%

\usepackage{mathtools}
\DeclarePairedDelimiter{\abs}{\lvert}{\rvert}
\usepackage{amsmath}
\usepackage{color}

\usepackage{natbib}

\theoremstyle{thmstyletwo}%

\theoremstyle{thmstylethree}%

\begin{document}

\title[
Self-organised magnon condensation
]{
Self-organised magnon condensation in quasi-1D edge-shared cuprates without external fields
}


\author*[1]{\fnm{Cli\`o Efthimia} \sur{Agrapidis}}\email{clio.agrapidis@fuw.edu.pl}

\author[2]{\fnm{Stefan-Ludwig} \sur{Drechsler}}\email{s.l.drechsler@ifw-dresden.de}

\author*[2,3]{\fnm{Satoshi} \sur{Nishimoto}}\email{s.nishimoto@ifw-dresden.de}

\affil[1]{Institute of Theoretical Physics, Faculty of Physics,
University of Warsaw, Pasteura 5, PL-02093 Warsaw, Poland}

\affil[2]{Institute for Theoretical Solid State Physics, IFW Dresden, 01069 Dresden, Germany}

\affil[3]{Department of Physics, Technical University Dresden, 01069 Dresden, Germany}


\abstract{
Multimagnon bound states were predicted nearly a century ago and have since been a key topic in condensed matter physics due to their intriguing quantum properties. However, their realization in natural materials remains elusive, especially in low-dimensional quantum magnets, where stabilizing them is particularly challenging due to the traditionally required extreme external magnetic fields. Therefore, we introduce a novel mechanism that enables the stabilization of multimagnon bound states in quasi-one-dimensional edge-shared cuprates. Our theoretical framework, supported by numerical simulations and experimental data, demonstrates that small antiferromagnetic interchain couplings act as effective internal magnetic fields, promoting a collinear antiferromagnetic order and enabling magnon condensation even at zero external field. This intrinsic stabilisation mechanism eliminates the need for high external fields, offering a platform that is more accessible for experimental realization. We validate this concept by applying it to representative materials such as Li$_2$CuO$_2$, Ca$_2$Y$_2$Cu$5$O$_{10}$, LiCuSbO$_4$, and PbCuSO$_4$(OH)$_2$. Beyond its experimental feasibility, this mechanism could drive advancements in magnon-based quantum computing, low-power spintronic devices, and high-speed magnonic circuits. Moreover, our findings reveal that small interchain and/or interlayer couplings can generally unlock previously overlooked magnetic phenomena, redefining the nature of magnetically ordered states and expanding the frontiers of quantum magnetism.
}


\maketitle

\section{Introduction}\label{sec:intro}

Understanding the dynamics of magnons in quantum magnetism is a captivating and evolving area of condensed matter physics. Magnons, the quasiparticles associated with spin waves in magnetically ordered systems, exhibit unique collective behaviors, including the formation of multimagnon bound states (MBSs). Over 90 years ago, Bethe~\cite{Bethe1931} predicted the existence of two-magnon bound states in one-dimensional (1D) quantum magnets, where ferromagnetic (FM) interactions bind two spin excitations together~\cite{Bethe1931, Wortis1963, Hanus1963}. Since then, the study of MBSs has become central to exploring the interplay between quantum interactions and collective excitations. Beyond the theoretical interest, MBSs have significant implications for quantum computing~\cite{Alicea2012}, spintronics~\cite{moore2010}, and emerging technologies like magnonics~\cite{Kruglyak2010}, where controlled manipulation of magnons is a key objective.

Advancements in quantum simulation have opened new pathways for probing MBSs. For instance, experiments with ultracold atoms in optical lattices have demonstrated magnon excitations and their binding into two-magnon states in quantum Heisenberg chains~\cite{Schaefer2020, Fukuhara2013}. Despite this progress, identifying MBSs in naturally occurring materials remains a critical goal. Pursuing this possibility, theoretical studies~\cite{Chubukov1991, Heidrich-Meisner2006, Vekua2007, Kecke2007, Hikihara2008, Sudan2009, Zhitomirsky2010, Sato2013} have extensively investigated the emergence of MBSs in frustrated ferromagnetic-antiferromagnetic (FM-AFM) $J_1$-$J_2$ chain systems, which are typical magnetic models for quasi-1D (Q1D) edge-shared cuprates~\cite{Kuzian2023}. These studies have demonstrated that MBSs can be induced by a magnetic field, and interestingly, near saturation, the types of MBSs vary with the ratio of $J_1$ to $J_2$, transitioning among two-magnon (nematic), three-magnon (triatic), four-magnon (quartic), and so on. Subsequent experimental efforts have aimed to identify MBSs, focusing on highly 1D materials such as LiVCuO$_4$, where signatures of nematic states have been observed~\cite{Svistov2011, Mourigal2012, Nawa2013, Buettgen2014, Orlova2017, Grafe2017}. Nevertheless, the stabilization of MBSs generally requires high external magnetic fields, making experimental observation challenging due to their fragility against thermal fluctuations and external perturbations~\cite{Ueda2009, PhysRevB.92.214415}. This challenge motivates the search for alternative stabilization mechanisms.

\begin{figure}
	\centering
	\includegraphics[width=0.8\linewidth]{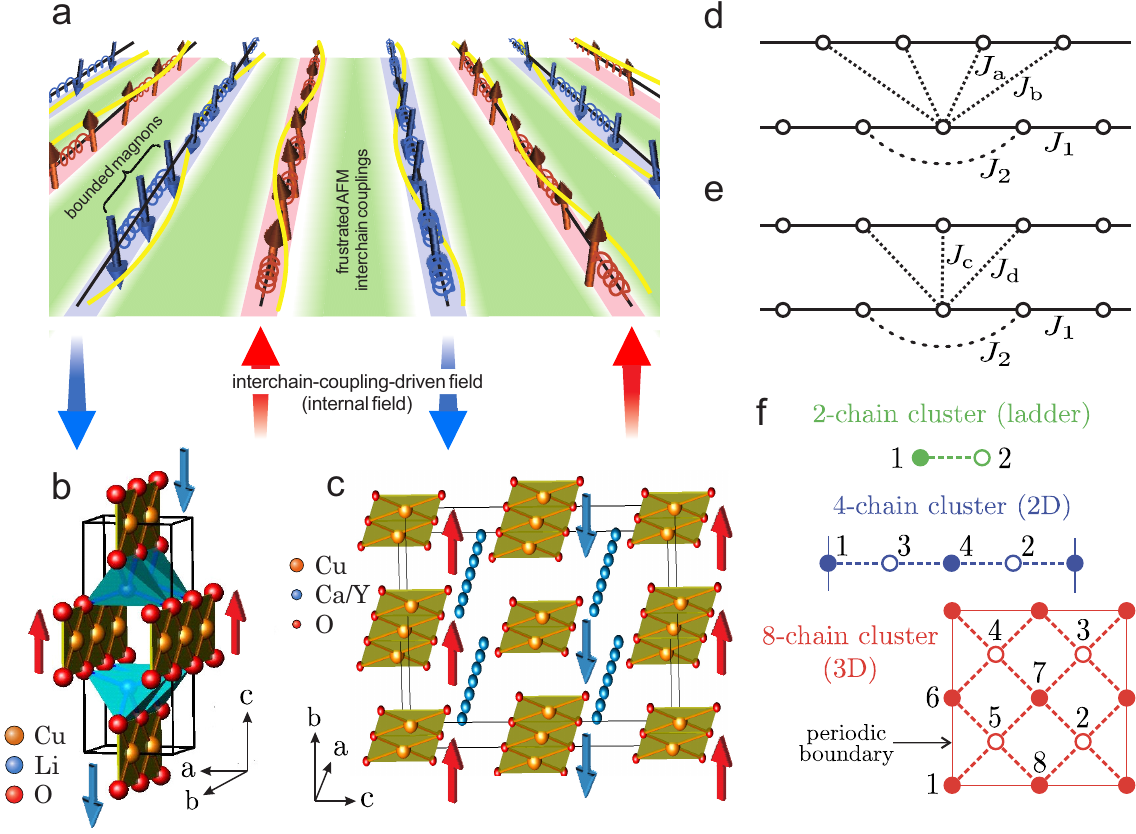}
	\caption{
		(a) Schematic illustration of the MBS driven by interchain coupling, corresponding to a triatic state with three magnons bound together. The yellow lines represent transverse spin fluctuations.
		Crystal structures of (b) Li$_2$CuO$_2$, showing two CuO$_2$ chains per unit cell along the b axis, and (c) Ca$_2$Y$_2$Cu$5$O$_{10}$, showing two CuO$_2$ chains per unit cell along the a axis. Red and blue arrows in (b) and (c) indicate an example of symmetry-broken magnetisation directions in the CAFO state. Structure of interchain couplings for compounds such as (d) Li$_2$CuO$_2$ and Ca$_2$Y$_2$Cu$5$O$_{10}$, and (e) LiCuSbO$_4$ and PbCuSO$_4$(OH)$_2$. (f) Schematic of the chain positions used in the numerical 2-chain, 4-chain, and 8-chain calculations, where the dashed lines denote the interchain coupling (see also Ref.~\cite{J_Phys_Conf_Ser_400_032069}).
	}\label{fig:lattice}
\end{figure}

In this study, we propose a novel mechanism for stabilizing MBSs in Q1D edge-shared cuprates, a material class known for its complex magnetic interactions. Our approach departs from conventional reliance on external magnetic fields. Instead, we demonstrate that small AFM interchain couplings can act as effective {\it internal fields}, inducing collinear antiferromagnetic order (CAFO) within individual chains. This order stabilizes robust MBSs through a self-organised mechanism. Unlike field-driven MBSs, which are often experimentally challenging to observe, this mechanism relies on intrinsic interchain interactions that are more accessible in real materials. This self-organised magnon condensation offers a promising new pathway for experimental realisation, shifting the paradigm of MBS stabilization. A conceptual picture of our MBS mechanism is illustrated in Fig.~\ref{fig:lattice}a.

The implications of this mechanism extend beyond MBSs in Q1D systems. Our results suggest that small interchain couplings can induce previously unobserved magnetic phenomena, providing fresh insights into the nature of magnetically ordered states. This perspective invites a reevaluation of magnetic systems across different dimensionalities, where subtle coupling interactions may unlock hidden magnetic behaviors. By demonstrating that interchain coupling-driven effects can mimic external fields, our findings open new avenues for exploring the interplay between quantum magnetism and material geometry.

In doing so, this work not only advances the understanding of MBSs in Q1D systems but also establishes a foundation for uncovering hidden magnetic behaviors in higher-dimensional systems. We begin by outlining the theoretical framework for MBS formation, emphasizing the role of AFM interchain couplings. Subsequently, we present detailed numerical results and experimental comparisons. We conclude with a discussion on broader implications and future research directions.

\section{Magnon bound states by external field}\label{sec:MBSext}

Prior to discussing our novel emergence mechanism of MBSs, it is necessary to outline the concept of MBSs and to explain how to theoretically calculate the magnon binding energy, which is one of the most direct physical quantities for detecting MBS. To illustrate the conventional MBS that emerges when spin-rotation symmetry is broken by an external field, we utilize the single FM-AFM $J_1$-$J_2$ chain. This analysis elucidates the key properties of MBSs, including the magnon binding energy as a function of the ratio between $J_1$ and $J_2$. These descriptions provide essential information for understanding the internal-field-induced MBSs discussed later.

\subsection{Magnon binding}\label{subsec:mb}

Magnons are collective excitations that arise from quantised spin wave perturbations in magnetically ordered systems, including ferromagnets, antiferromagnets and helimagnets. They follow Bose-Einstein statistics, classifying them as bosons. The presence of a magnon signifies a flip in the spin orientation of an electron relative to its ordered (ground) state. It is notable that in environments with strong magnon-magnon interactions and enhanced magnetic correlations, which are typically observed in low-dimensional systems prone to magnetic frustration, magnons exhibit unique behaviors, including the ability to pair and form MBSs. Such pairing results in a composite excitation known as a bimagnon, analogous to Cooper pairs in superconductors. 

The quantum phase referred to as a spin nematic state, i.e., a two-magnon bound state, is characterised by a predominance of collective bimagnon excitations. It features a preferred orientation of the paired magnons without spatial confinement, breaking spin-rotational symmetry while retaining translational symmetry. This phenomenon is typically realised in conditions where spins are polarised, for instance, by an external magnetic field. Furthermore, under specific conditions, higher-order bound states involving more than two magnons, such as `triatic' or `quartic' states, may form. These states arise from the intricate interactions of three or four magnons, respectively.

An effective theoretical approach to identifying MBS is to calculate the binding energy of magnons, which is crucial for understanding their interactions and properties. This energy quantifies the cost of dissociating a bound state into individual magnons. This method is in analogy with the binding energy of Cooper pairs in superconductivity~\cite{Chakravarty1991}. When a MBS forms, it is associated with a specific binding energy. To calculate this for a $p$-MBS (where $p$ is the number of magnons, e.g. 2 for nematic, 3 for triatic, 4 for quartic state) at a given net magnetisation $M$ (assuming conserved $S^z$), the formula is
\begin{align}
	E_b(M,p)\!=\left[E_0(S^z\!=\!M\!-\!1)\!-\!E_0(S^z\!=\!M)\right]\!
	-\!\frac{1}{p}\left[E_0(S^z\!=\!M\!-\!p)\!-\!E_0(S^z\!=\!M)\right],
	\label{eq:be}
\end{align}
where the ground state energy for a total spin $S^z=S$ is denoted by $E_0(S^z=S)$. A positive value of the binding energy in the bulk limit indicates the formation of a bound state, which is an important indicator of the presence of MBS. However, unlike for the electron binding energy in superconductivity, the number of bound magnons is not known in advance, so it is necessary to calculate this quantity for different values of $p$. If the largest value of $E_b(M,p)$ ($>0$) is given by $p=p_{\rm max}$, it can be concluded that the $p_{\rm max}$ MBS is more energetically favorable than both the individual magnons and other $p$ MBSs. A schematic illustration of the nematic case ($p=2$) is shown in Fig.~\ref{fig:singlechain}a. The magnon binding energy is observed in experiments as a kind of excitation gap. The following subsection illustrates the calculation of the magnon binding energy for the FM-AFM $J_1$-$J_2$ chain.

\subsection{1D frustrated chain at high fields}\label{subsec:mb1dhf}

\begin{figure}
	\centering
	\includegraphics[width=1.0\linewidth]{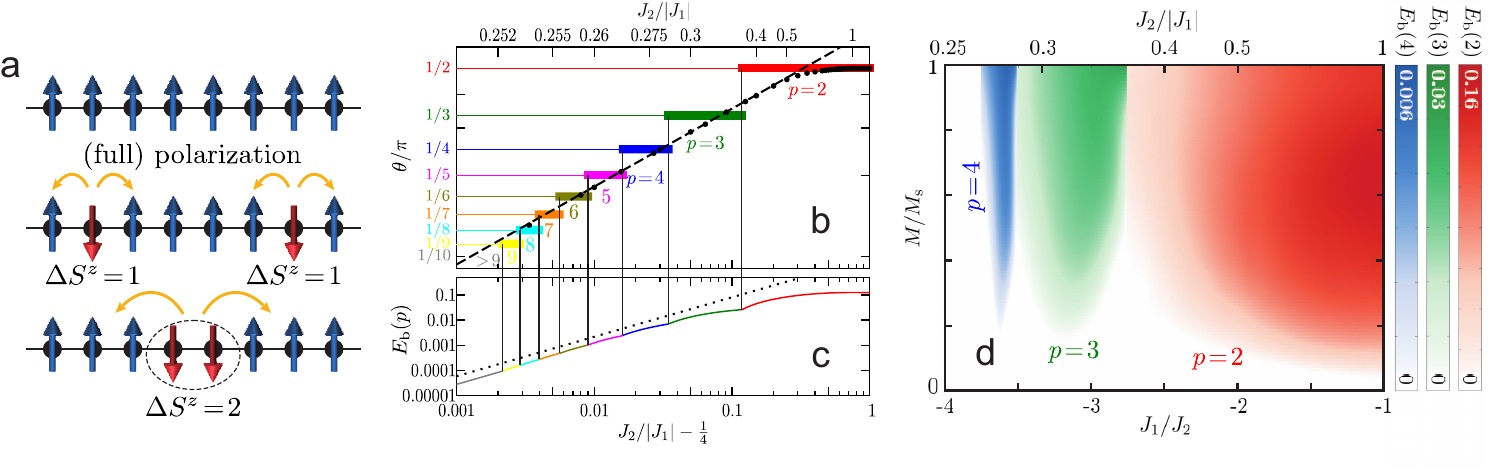}
	\caption{
		(a) Schematic illustrations of (top) the fully polarised state, (middle) the state with two flipped spins far apart -- individual magnons --, and (bottom) the state with two adjacent flipped spins -- bounded magnons --.
		(b) Propagation number $\theta$ (black circles) of the $J_1$-$J_2$ chain at zero magnetic field as a function of $J_2/\abs{J_1}$, and the relationship between the number of bound magnons $p$ in the MES at the saturation field. The propagation number is fitted by $\theta/\pi=0.69(J_2/\abs{J_1})^{0.29}$ near $J_2/\abs{J_1}=1/4$.
		(c) Magnon binding energy $E_b(M_{\rm s}, p)$ corresponding to $p$ in the MES at the saturation field. The dashed line denotes $E_b(M_{\rm s}, p) \propto (J_2/\abs{J_1})^{\pi/2}$.
		(d) Ground state phase diagram for MBS as a function of $J_1/J_2$ (and $J_2/\abs{J_1}$) and $M/M_{\rm s}$, with the magnon binding energy $E_b(M, p)$ depicted in a colour density plot corresponding to each phase.
	}\label{fig:singlechain}
\end{figure}

One of the most promising systems exhibiting MBS is the 1D Heisenberg chain with frustrated FM nearest neighbor (NN) and AFM next nearest neighbor (NNN) interactions, namely the FM-AFM $J_1$-$J_2$ chain, under an external magnetic field. The Hamiltonian for this system is expressed as follows
\begin{align}
	{\cal H}_\alpha=J_1\sum_i\vec{S}_{\alpha,i}\cdot\vec{S}_{\alpha,i+1}+J_2\sum_i\vec{S}_{\alpha,i}\cdot\vec{S}_{\alpha,i+2}-h\sum_i S^z_{\alpha,i},
	\label{eq:1Dham}
\end{align}
where $\mbox{\boldmath $S$}_{\alpha,i}=(S^x_{\alpha,i},S^y_{\alpha,i},S^z_{\alpha, i})$ represent the spin $\frac{1}{2}$ operators (equivalent to the Pauli matrices) at the $i$-th site on the chain $\alpha$ (the index $\alpha$ is used later for coupled chains), and $S^\pm_{\alpha,i}=S^x_{\alpha,i} \pm iS^y_{\alpha,i}$. The symbols $J_1$ ($<0$) and $J_2$ ($>0$) denote the FM NN and AFM NNN interactions respectively. The external magnetic field is controlled by $h$. This system is often used to describe the magnetic properties of quasi-1D edge-shared cuprates, where the NN interaction is typically FM due to the direct overlap of the copper and oxygen orbitals along the edge~\cite{Kuzian2023}.

The ground state of the system~\eqref{eq:1Dham} at zero field ($h=0$) has been studied extensively~\cite{Tonegawa1989,Bursill1995,Nersesyan1998,Itoi2001,Sirker2011,Furukawa2012,Agrapidis2019}. There are two distinct phases: the FM phase at $J_2/\abs{J_1}<1/4$ and the incommensurate spiral phase at $J_2/\abs{J_1}>1/4$. Throughout the spiral phase, the incommensurate spin-spin correlations are short-range~\cite{Nersesyan1998,Sirker2011}; instead, a spontaneous NN-FM dimerization order accompanied by translational symmetry breaking occurs due to the order-by-disorder mechanism~\cite{Furukawa2012,Agrapidis2019}. Consequently, the system exhibits a form of valence bond solid (VBS) state with spin singlet formations between third-neighbor sites~\cite{Agrapidis2019}. By considering the ferromagnetically dimerised spin-1/2 pair as a spin-1 site, this VBS state can also be recognised as an Affleck-Kennedy-Lieb-Tasaki~\cite{Affleck1988} or Haldane~\cite{Haldane1983-1,Haldane1983-2} state with topological order protected by global $Z_2 \times Z_2$ symmetry~\cite{Oshikawa1992}.

The theoretical phase diagram at finite magnetic fields has attracted considerable attention in the context of MBSs~\cite{Chubukov1991, Heidrich-Meisner2006, Vekua2007, Kecke2007, Hikihara2008, Sudan2009}. When an external magnetic field is applied to the spiral phase, a vector chiral phase appears. This phase is characterised by the breaking of the discrete parity symmetry ($Z_2$ symmetry breaking) accompanied by a spontaneous $S^z$ spin current circulation~\cite{Hikihara2008}. The longitudinal vector chiral order parameter, $\kappa_l^{(n)} = \langle (\mbox{\boldmath $S$}_l \times \mbox{\boldmath $S$}_{l+n})^z \rangle$, acquires a non-zero value in this phase. As the magnetic field strength increases, an MBS appears. The number of bound magnons depends on the ratio of $J_2$ to $\abs{J_1}$. By calculating the binding energy $E_{\rm b}(M,p)$, expressed in eq.~\ref{eq:be}, the number of bound magnons can be determined. Fig.~\ref{fig:singlechain}b shows the propagation number $\theta$ at zero field and the number of bound magnons $p$ near full saturation ($M/M_{\rm s}=1$) as a function of $J_2/\abs{J_1}-1/4$. The corresponding binding energy in the thermodynamic limit is plotted in Fig.~\ref{fig:singlechain}c. Approaching the FM phase transition at $J_2/\abs{J_1}=1/4$, the number of bound magnons increases while the binding energy decreases sharply, indicating that larger bound magnon sizes are more susceptible to external perturbations. The conjectured relationship between $\theta$ and $p$, $1/p > \theta/\pi > 1/(p+1)$~\cite{Sudan2009}, holds over a wide range of $J_2/\abs{J_1}$ values, except near $J_2/\abs{J_1}=1/4$. Furthermore, Fig.~\ref{fig:singlechain}d shows the distribution of $E_{\rm b}(M, p)$ as a function of $J_1/J_2$ and $M/M_{\rm s}$. We observe that the binding energy peaks slightly below the saturation magnetisation $M/M_{\rm s}=1$, suggesting that sufficient breaking of the spin-rotation symmetry allows bound magnons to move parallel or antiparallel to the magnetic field, thereby gaining propagation energy.

\subsection{Experimental Challenges}\label{subsec:exp}

The experimental observation of MBS in Q1D systems presents significant difficulties. MBS can be detected by a variety of experimental methods, including the temperature dependence~\cite{Sato2009, Momoi2024, Grafe2017} and field dependence~\cite{Zhitomirsky2010, Orlova2017} of the nuclear magnetic resonance (NMR) relaxation rate, field dependent wave vector measurements by neutron scattering experiments~\cite{Vekua2007, Kecke2007, Hikihara2008, Sudan2009, Affleck1988, Mourigal2012} and field dependent magnetisation~\cite{Zhitomirsky2010, Svistov2011}. Fig.~\ref{fig:singlechain}d suggests that the MBS phase spans a wide range of magnetisation. However, the substantial quantum fluctuations in Q1D systems result in a very gradual increase in magnetisation at low fields. Thus, high magnetic fields are required to achieve sufficient magnetisation to stabilize MBS in these systems. Furthermore, while the binding energy of magnons in an isolated $J_1$-$J_2$ chain is already small, it decreases sharply with increasing AFM interchain interactions. As a result, the MBS is extremely sensitive to these AFM interchain interactions~\cite{Ueda2009,PhysRevB.92.214415}. Therefore, experimental setups often require conditions such as low temperatures and high magnetic fields to stabilize and observe MBS. This results in keeping the practical observation of external field induced MBS a major challenge.

\section{Magnon bound states by internal field}\label{MBSin}

As previously discussed, the observation of MBS in Q1D systems has traditionally been a matter of contention, primarily due to the challenges in experimental detection. However, the novel mechanism proposed herein facilitates magnon binding via AFM interchain coupling, which can mimic the effect of an actual external magnetic field on each chain, thereby eliminating the need for an actual external magnetic field. This advancement presents a significant advantage for experimental observations. This section elucidates the underlying mechanism and demonstrates that MBS can indeed emerge within realistic parameter regimes.

\subsection{Internal field from mean-field approximation of interchain couplings}\label{subsecMF}

In Q1D magnetic systems, the interplay between interchain and intrachain couplings plays a pivotal role in determining the magnetic ground state. Notably, when interchain couplings are significantly weaker than their intrachain counterparts, the interchain interactions can be effectively approximated through a mean-field approach. This approximation allows for the conceptualization of the system as a collection of independent 1D chains influenced by an effective mean field. This concept was initially validated for isotropic, i.e., XXX, Heisenberg chains linked by perpendicular interchain couplings $J_\perp$, described by the Hamiltonian ${\cal H}=\sum_{i,j}\mbox{\boldmath $S$}_{i,j}\cdot\mbox{\boldmath $S$}_{i,j+1}+J_\perp\sum_{i,j}\mbox{\boldmath $S$}_{i,j}\cdot\mbox{\boldmath $S$}_{i+1,j}$ with $J_\perp>0$. In the case of $J_\perp=0$, the system lacks magnetic order and is classified as a Tomonaga-Luttinger liquid at zero temperature. However, the spin-spin correlation functions decay in a power law manner, indicating the presence of quasi-long-range order. Hence, if finite $J_\perp$ is switched on, the system exhibits N\'eel long-range order~\cite{Affleck1994}.

Assuming the order to be oriented along the $z$-direction in spin space, the Hamiltonian transforms into an effective single-chain problem described by ${\cal H}_{\rm eff}=J\sum_i\mbox{\boldmath $S$}_i\cdot\mbox{\boldmath $S$}_{i+1}-h_{\rm stag}\sum_i(-1)^iS^z_i+{\rm const.}$, where $h_{\rm stag}$ denotes the staggered field. This field, acting as an {\it effective internal force}, encapsulates the collective effect of the interactions with adjacent chains, inherent to the structure of the system. It elucidates the formation of N\'eel order in 2D square and 3D cubic lattices as a dimensional extension from a 1D chain. The configuration of this effective internal field depends on the structural arrangement of magnetic interactions within the system. For example, it can exhibit incommensurate oscillations if the net magnetisation of the system is nonzero~\cite{Clio2019,Nikitin2021}.

What are the potential forms of internal fields in Q1D FM-AFM $J_1$-$J_2$ chain systems? The Hamiltonian, incorporating interchain coupling, can be expressed as
\begin{align}
	\nonumber
	{\cal H}=&\underbrace{J_1\sum_{\alpha,i}[\Delta_1(S^+_{\alpha,i}S^-_{\alpha,i+1}+S^-_{\alpha,i}S^+_{\alpha,i+1})+S^z_{\alpha,i}S^z_{\alpha,i+1}]+J_2\sum_{\alpha,i}\vec{S}_{\alpha,i}\cdot\vec{S}_{\alpha,i+2}}_{\text{intrachain couplings}}\\
	&+\underbrace{\sum_{\substack{\alpha\beta,ij \\ \gamma=a,b\, {\rm or}\, c,d}}J_\gamma[\Delta_\gamma(S^+_{\alpha,i}S^-_{\beta,j}+S^-_{\alpha,i}S^+_{\beta,j})+S^z_{\alpha,i}S^z_{\beta,j}]}_{\text{interchain couplings}}
	\label{eq:ham}
\end{align}
where $J_\gamma$ ($\gamma={\rm a,b\, or\, c,d}$) denotes the interchain coupling, the common structures of which are illustrated in Fig.~\ref{fig:lattice}d,e. For example, compounds corresponding to Fig.~\ref{fig:lattice}d include Li$_2$CuO$_2$ and Ca$_2$Y$_2$Cu$5$O$_{10}$, while those corresponding to Fig.~\ref{fig:lattice}e include LiCuSbO$_4$, PbCuSO$_4$(OH)$_2$, and LiVCuO$_4$. In the direction perpendicular to the chain, they have a bipartite structure. If one could neglect the effects of interchain coupling, each chain would exhibit a spiral state with incommensurate short-ranged spin-spin correlations if $J_2/\abs{J_1}>1/4$. The pitch angle $\phi$ of this spiral, dependent of $J_2/\abs{J_1}$, ranges between 0 and $\pi/2$, and $\phi=\cos^{-1}[-J_1/(4J_2)]$ in the classical limit. When the interchain couplings are AFM, it is known that the pitch angle decreases as their coupling strength increases, potentially leading to FM polarisation of each chain when $\phi$ reaches zero~\cite{EPL2012}. This indicates that the ratio $J_2/\abs{J_1}$ is effectively diminished by AFM interchain coupling, reaching an FM critical point $1/4$ at a certain threshold. Conversely, it can also be said that AFM interchain coupling shifts the FM critical point towards higher $J_2/\abs{J_1}$ values. The effective FM critical point can be analytically calculated, yielding for the cases represented in Fig.~\ref{fig:lattice}d and \ref{fig:lattice}e, 
\begin{align}
	\left(\frac{J_2}{\abs{J_1}}\right)_{\rm FM,c}=\frac{1}{4}+\frac{1}{8}\frac{J_a+9J_b}{\abs{J_1}}
	\label{eq:JFMc1}
\end{align}
and
\begin{align}
	\left(\frac{J_2}{\abs{J_1}}\right)_{\rm FM,c}=\frac{1}{4}+\frac{J_d}{2\abs{J_1}},
	\label{eq:JFMc2}
\end{align}
respectively. Note that this analysis is for a 2D system, where each chain has two adjacent chains. For a 3D system, where each chain has four adjacent chains, these equations are valid with the replacements $J_a \to 2J_a$, $J_b \to 2J_b$, and $J_d \to 2J_d$~\cite{EPL2012}.

If the magnitude of the AFM interchain coupling is sufficiently large to induce FM ordering in each chain, the spins in adjacent chains will be polarised in opposite directions. This scenario represents a phase transition from a spiral state to a CAFO, governed by the AFM interchain coupling. By conceptualizing the AFM interchain coupling as an effective internal field, each chain can be viewed as an FM-AFM $J_1$-$J_2$ chain under a uniform field, with the field direction being opposite in adjacent chains. If this mapping holds true, the effective Hamiltonian for a single chain should be expressible in a form similar to Equation~\eqref{eq:1Dham}. In other words, if the internal field induced by the AFM interchain coupling can break the spin symmetry of each chain, leading to a finite magnetisation, it might be possible to realise MBSs without the need for an external magnetic field. In the following subsections, we will numerically verify the formation of CAFO as the AFM interchain coupling is increased, and the realisation of MBSs.

\subsection{Magnetisation in self-organised collinear antiferromagnetic ordering}\label{subsecSOCO}

\begin{figure}
	\centering
	\includegraphics[width=1.0\linewidth]{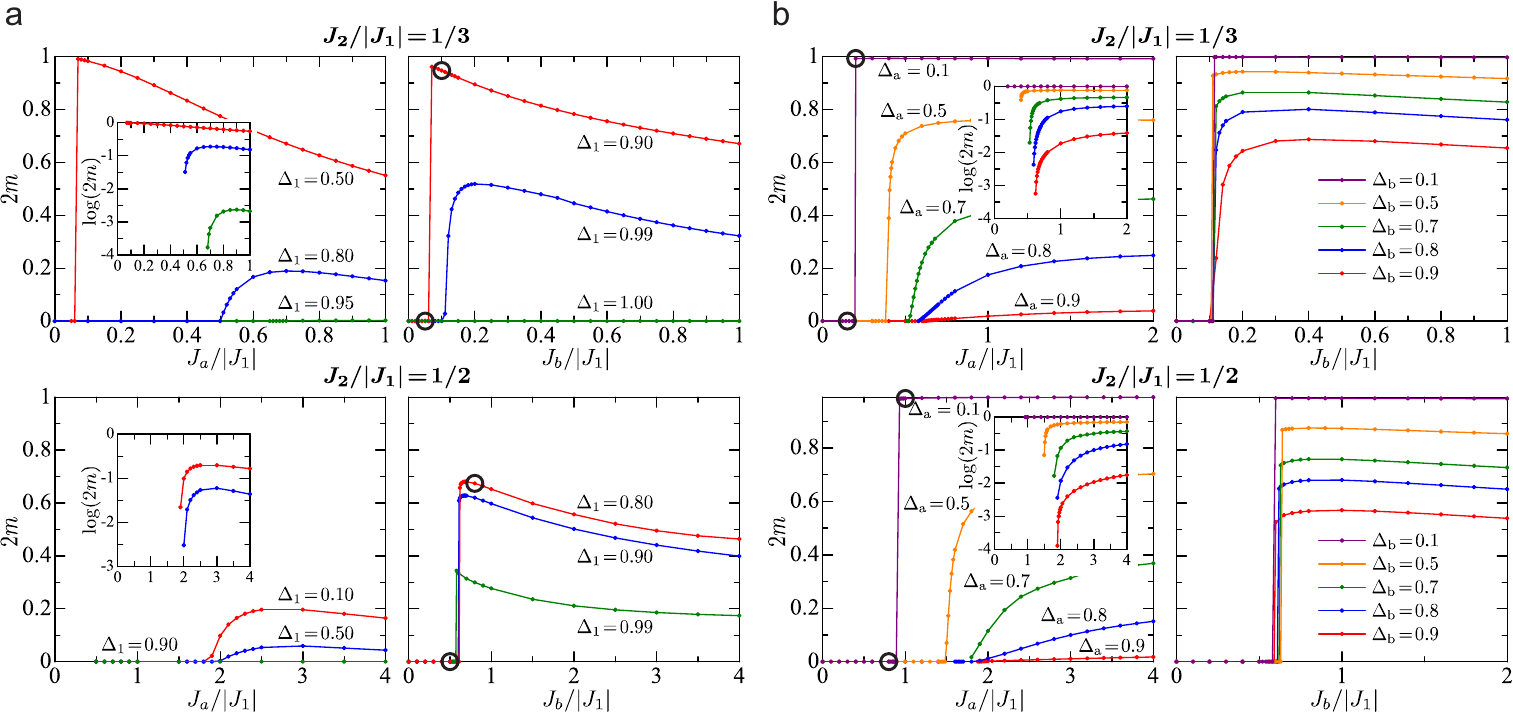}
	\caption{Magnetisation per site in the thermodynamic limit ($l_x \to \infty$) for various strengths of the XXZ anisotropy, evaluated at representative ratios $J_2/\abs{J_1}=1/3$ and $1/2$. The magnetisation is presented as a function of the interchain couplings $J_{\rm a}$ and $J_{\rm b}$. Panels (a) and (b) illustrate the effects of introducing XXZ anisotropy on inchain and interchain couplings, respectively. The parameters used in the calculations of MES are marked with circles (see Fig.~\ref{fig:mes}). 
	}
	\label{fig:mag_Jic}
\end{figure}

Here, we demonstrate the formation of CAFO driven by AFM interchain interaction in the coupled FM-AFM $J_1$-$J_2$ chains. To align the magnetisation direction along the z-axis, we introduce XXZ spin anisotropy into either the intrachain or interchain couplings, a feature known to exist, albeit moderately, in real materials. The magnetisation per site, serving as an order parameter for the CAFO, is defined by:
\begin{align}
	m=\frac{1}{l_x l_y}\left\lvert \sum_{\alpha,j}(-1)^j \langle S^z_{\alpha,j}\rangle \right\lvert
\end{align}
Given the bipartite lattice configuration perpendicular to the chains, we segregate the chain indices $j$ into even and odd numbers for each sublattice. For a comprehensive exploration of parameter dependencies, we confine our system to 2-chain cluster ($l_y=2$), a scenario that maximizes quantum fluctuations, thereby posing a challenging environment for the establishment of the CAFO. The formation of a CAFO under such stringent conditions suggests that more stable CAFO structures would naturally arise in systems with reduced quantum fluctuations, such as in two- or three-dimensional lattices.  Indeed, we have confirmed a more pronounced CAFO stabilisation for 4-chain ($l_y=4$) and 8-chain ($l_y=8$) clusters illustrated in Fig.~\ref{fig:lattice}f in comparison to the $l_y=2$ case.

Using the density-matrix renormalization group (DMRG) method, we calculate the magnetisation per site in the thermodynamic limit ($l_x \to \infty$) for various strengths of the XXZ anisotropy and representative values of the ratio $J_2/\abs{J_1}=1/3$ and $1/2$. Details on our DMRG calculations are given in Methods. The results are plotted as a function of the interchain couplings $J_{\rm a}$ and $J_{\rm b}$ in Fig.~\ref{fig:mag_Jic}. In Fig.~\ref{fig:mag_Jic}a, the XXZ anisotropy ($\Delta_1 < 1$) is introduced in the intrachain coupling, while in Fig.~\ref{fig:mag_Jic}b, the XXZ anisotropy ($\Delta_{\rm a,b} < 1$) is included in the interchain coupling. In both cases, a critical interchain coupling exists for all combinations of $J_2/\abs{J_1}$ and a kind of the interchain couplings, beyond which the system transitions to the CAFO phase. When $J_1$-$J_2$ chains in the spiral phase are coupled by frustrated AFM interchain couplings, increasing the interchain coupling reduces the propagation number. Once this number reaches zero, the system transitions to the CAFO phase~\cite{EPL2012}. Therefore, comparing the results for $J_2/\abs{J_1}=1/3$ and $J_2/\abs{J_1}=1/2$, the interchain coupling required to induce the transition to the CAFO phase is smaller for $J_2/\abs{J_1}=1/3$, which originally has a smaller propagation number. Furthermore, as inferred from Eq.~\ref{eq:JFMc1}, the effect of $J_{\rm b}$ is generally larger than that of $J_{\rm a}$. For instance, for $J_2/\abs{J_1}=1/3$, a slight XXZ anisotropy in the intrachain coupling leads to a significant magnetisation when the interchain coupling is $J_{\rm b}$. This scenario corresponds to the case of Li$_2$CuO$_2$ discussed in Sec.~\ref{subsec:li2cuo2}.

Thus, it was found that each $J_1$-$J_2$ chain becomes magnetised by the AFM interchain coupling as if an external uniform field had been applied. In the following subsection, we will verify whether MBS indeed emerges in this magnetised state.

\subsection{Magnon excitation spectrum}\label{subsecMES}

\begin{figure}
	\centering
	\includegraphics[width=1.0\linewidth]{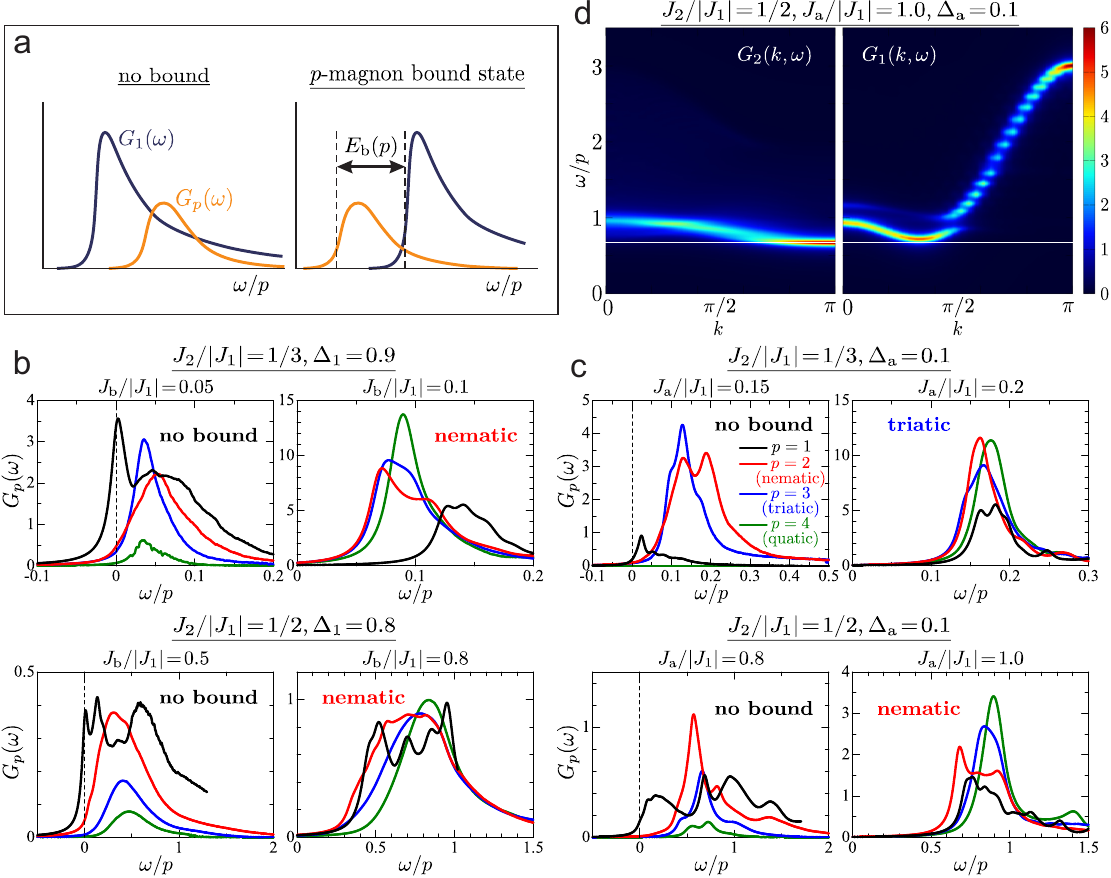}
	\caption{(a) Schematic illustration distinguishing a MBS from a non-MBS within the MES framework. (b,c) DDMRG results for the MES $G_p(\omega)$ at various values of $p=1$, $2$, $3$, and $4$, for representative ratios $J_2/\abs{J_1}=1/3$ and $1/2$, with the XXZ anisotropy applied to (b) intrachain and (c) interchain couplings. (d) Angle-resolved MES $G_1(k,\omega)$ and $G_2(k,\omega)$ for $J_2/\abs{J_1}=1/2$, $J_{\rm a}/\abs{J_1}=1.0$, and $\Delta_{\rm a}=0.1$.
	}
	\label{fig:mes}
\end{figure}

In real materials, the behavior of coupled $J_1$-$J_2$ chains can be particularly fascinating when considering MBSs. A straightforward approach to detect these states is to calculate the binding energy of magnons, $E_{\rm b}(M,p)$, as done in Sec.~\ref{subsec:mb1dhf} using Eq.~(\ref{eq:be}). However, this approach faces a problem for coupled chains due to interchain exchange interactions, which lead to the non-conservation of the total $S^z$ within each chain. Consequently, Eq.~(\ref{eq:be}) cannot be directly applied. To circumvent this issue, we investigate the presence of MBSs by calculating the magnon excitation spectrum (MES) for scenarios where $p$ spins are flipped simultaneously. The MES for $p$-magnon flips is defined as:
\begin{align}
	G_p(\omega)=\sum_{n>0}\lvert\langle n\lvert\prod_{i=r}^{r+p-1}S^-_i\lvert0\rangle\lvert^2\delta(\omega-E_n+E_0).
	\label{eq:mes}
\end{align}
This function is crucial for identifying MBSs. It has been established that when $p$ magnons form a bound state, the average size of these magnons is effectively $p-1$~\cite{Kecke2007}. This implies that the magnons are tightly bound and form the lowest-lying excitations in the energy spectra. Therefore, computing MES by flipping $p$ contiguous spins is a reasonable approach to explore the dynamics and interactions of MBSs in coupled chains. This methodology provides a deeper understanding of quantum magnetic phenomena and extends our insight into the collective excitations in complex spin systems.

To determine the presence of MBSs, one must first compute the MES for $p=1$, denoted as $G_1(\omega)$, and for $p>1$, denoted as $G_{p(>1)}(\omega)$. As illustrated in the left panel of Fig.~\ref{fig:mes}a, if the onset of $G_1(\omega)$ occurs at a lower energy than that of $G_{p(>1)}(\omega)$, this suggests that there is no binding among the magnons. Conversely, as shown in the right panel of Fig.~\ref{fig:mes}a, if the onset of $G_{p(>1)}(\omega)$ is at a lower energy than that of $G_1(\omega)$, it indicates that $p$ magnons are bound. In this scenario, the difference between the onsets of $G_{p(>1)}(\omega)$ and $G_1(\omega)$ corresponds to the magnon binding energy $E_{\rm b}(M,p)$. To accurately identify the state with the lowest energy onset, one needs to systematically vary $p$ through values such as $2$, $3$, $4$, and so on, and observe which $G_{p(>1)}(\omega)$ yields the minimum onset energy. This procedure enables the identification of the optimal number of bound magnons and provides a quantitative measure of the magnon binding energy. It is noteworthy that when MBSs are not formed, the system is essentially gapless. Consequently, the lowest excitation of $G_1(\omega)$, corresponding to inelastic neutron scattering (INS), is at zero energy.

We then investigate whether MBSs truly form when a CAFO is established by the internal field, as demonstrated in Sec.~\ref{subsecSOCO}. In the CAFO, each $J_1$-$J_2$ chain orders ferromagnetically, with neighboring $J_1$-$J_2$ chains having their magnetisations aligned anti-parallel along the $z$-axis. Here, we compute the MES for the $J_1$-$J_2$ chains with magnetisation pointing in the positive $z$-direction using Eq.~(\ref{eq:mes}). We note that an equivalent result is obtained for chains with magnetisation in the negative $z$-direction by replacing $S^-_i$ with $S^+_i$ in Eq.~(\ref{eq:mes}).

The MES computed using dynamical DMRG (DDMRG) method for various parameters is shown in Figs.~\ref{fig:mes}b and \ref{fig:mes}c. Details on our DDMRG calculations are given in Methods. The parameters used in those calculations are marked with circles in Fig.~\ref{fig:mag_Jic}. To clarify the impact of CAFO on the MES, we compare the MES before and after the critical interchain coupling. In Fig.~\ref{fig:mes}b, where $J_1$ is modified to include XXZ anisotropy, we observe that for $J_2/\abs{J_1}=1/3$ and $\Delta_1 = 0.9$, no magnon binding is present at $J_{\rm b} = 0.05$, as the lowest excitation is provided by $G_1(\omega)$. However, when $J_{\rm b} = 0.1$, leading to the formation of CAFO, $G_2(\omega)$ provides the lowest excitation, indicating that the ground state is a nematic state. Remarkably, the onset of $G_3(\omega)$ is very close to that of $G_2(\omega)$, suggesting that a triatic state exists very near the ground state. For $J_2/\abs{J_1}=1/2$ and $\Delta_1 = 0.8$, the MES transitions from $G_1(\omega)$ to $G_2(\omega)$ as the lowest excitation when CAFO forms.

Fig.~\ref{fig:mes}c shows the MES when XXZ anisotropy is introduced in the interchain coupling. Essentially, similar results are obtained as for XXZ anisotropy applied to $J_1$. For $J_2/\abs{J_1}=1/3$ and $\Delta_{\rm a}=0.1$, the ground state becomes a triatic state upon the formation of CAFO. For $J_2/\abs{J_1}=1/2$ and $\Delta_{\rm a}=0.1$, the ground state becomes a nematic state upon CAFO. The value of $p$ for the ground state generally corresponds to the number of magnon bindings stabilised by an external field in a single $J_1$-$J_2$ chain for a given $J_2/\abs{J_1}$. Given that $J_2/\abs{J_1}=1/3$ is very close to the boundary between $p=2$ and $p=3$ at $J_2/\abs{J_1}=0.367615$ in the high-field phase diagram of a single $J_1$-$J_2$ chain (see Figs.~\ref{fig:singlechain}b, c, d), the MES shows $G_2(\omega)$ and $G_3(\omega)$ onsets close to each other, almost degenerate at the ground state. However, in any case, the presence of CAFO results in excitations significantly lower than those given by $G_1(\omega)$.

In comparison with the properties of conventional MBSs, it would also be useful to examine the total momentum of the multi-magnon modes within the MBS formed by the internal field. To confirm this, we compute the angle-resolved MES, which can be defined as follows:
\begin{align}
	\nonumber
	G_p(k,\omega)=&\sum_{n>0}\lvert\langle n\lvert\sum_j\left(\prod_{i=j}^{j+p-1}S^-_i\right)\exp(ikr_j)\lvert0\rangle\lvert^2\\
	&\times \delta(\omega-E_n+E_0)
	\label{eq:mes_k}
\end{align}
The results of calculating $G_p(k,\omega)$ using DDMRG for $G_1(\omega)$ and $G_2(\omega)$ corresponding to $J_2/\abs{J_1}=1/2$, $J_{\rm a}/\abs{J_1}=1.0$, $\Delta_{\rm a}=0.1$ are presented in Fig.~\ref{fig:mes}d. Essentially, these results demonstrate that our novel MBS exhibits properties akin to those emerging in an isolated $J_1$-$J_2$ chain under an external magnetic field~\cite{Chubukov1991,Heidrich-Meisner2006,Vekua2007,Kecke2007,Hikihara2008,Sudan2009}. Specifically, the two-magnon excitation $G_2(k,\omega)$ shows a minimum at momentum $k=\pi$. Conversely, $G_1(k,\omega)$, which corresponds primarily to the transverse component of the dynamical spin structure factor $S^{xx}(k,\omega)$ (or $S^{yy}(k,\omega)$), exhibits a minimum around $k=0.332\pi$, reflecting the incommensurate correlations of the isolated $J_1$-$J_2$ chain in the absence of a magnetic field. Given the nematic ground state, the lowest bound of $G_2(k,\omega)$ is located lower than that of $G_1(k,\omega)$, consistent with the $G_1(\omega)$ and $G_2(\omega)$ shown in Fig.~\ref{fig:mes}c.

\section{Realisation of zero-field MBS in Q1D cuprates}\label{sec:realisation}

Finally, we consider four materials candidates for the potential formation of MBSs based on the mechanism proposed above.

\begin{figure}
	\centering
	\includegraphics[width=0.9\linewidth]{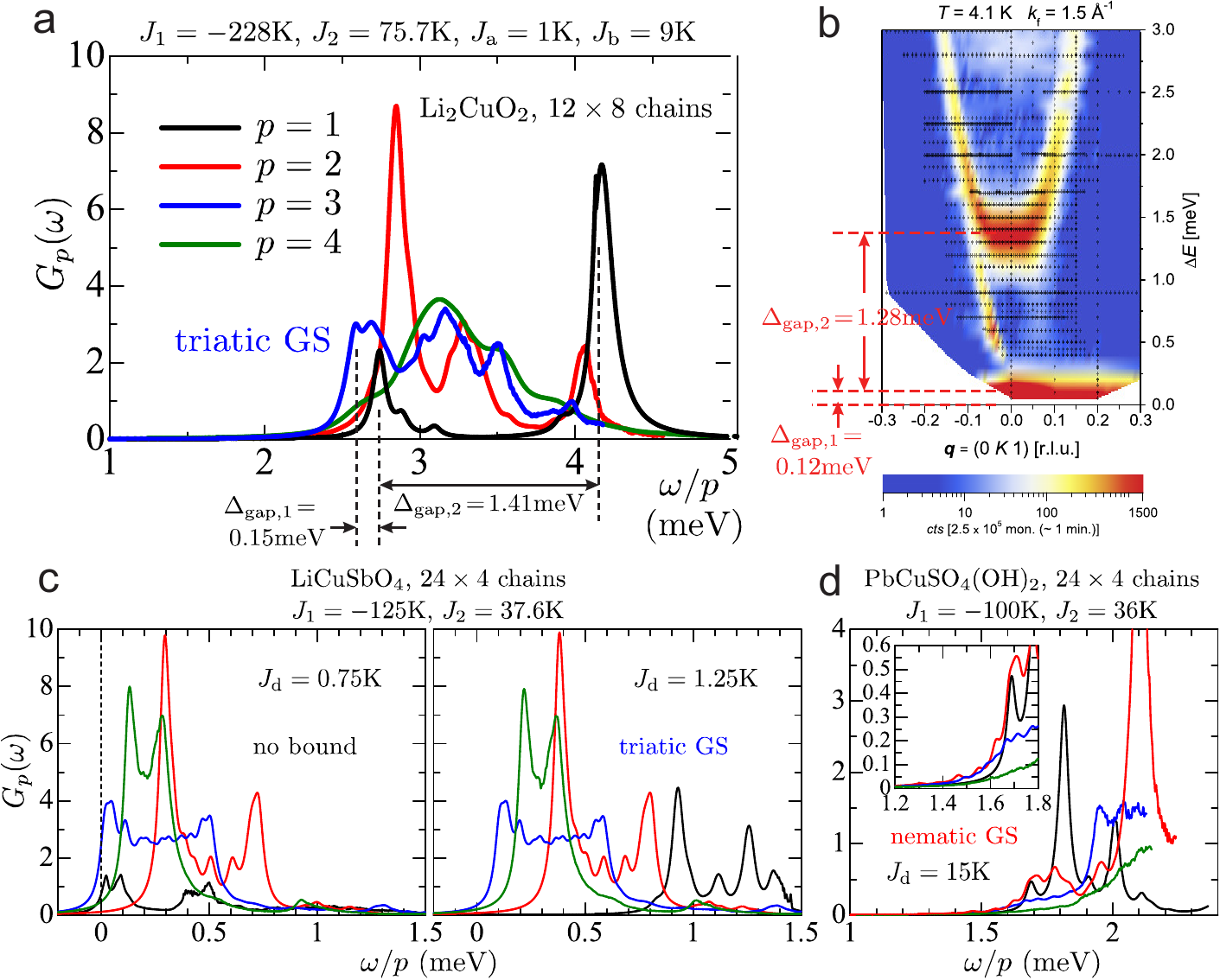}
	\caption{(a) DDMRG results for $G_p(\omega)$ ($p=1$ to $4$) using Li$_2$CuO$_2$ parameter set with $12 \times 8$-chain cluster.	
	(b) Intensity map of the INS data extracted from Ref.~\cite{Lorenz2009}.
	(c,d) DDMRG results for $G_p(\omega)$ ($p=1$ to $4$) using the intrachain couplings for LiCuSbO$_4$ and PbCuSO$_4$(OH)$_2$ with $24 \times 4$-chain cluster. The interchain AFM couplings are set to be tunable parameter.
	}\label{fig:MESmaterials}
\end{figure}

\subsection{Li$_2$CuO$_2$}\label{subsec:li2cuo2}

The most promising material for the application of our MBS emergence mechanism is Li$_2$CuO$_2$. The magnetic couplings for this material are estimated to be $J_1=-228$ K, $J_2=75.7$ K, $J_{\rm a}=1$ K, and $J_{\rm b}=9$ K~\cite{Lorenz2009}. Also, paramagnetic and AFM resonance experiments have observed easy-axis anisotropy along the a-axis~\cite{Ohta1993}, which is estimated to be approximately $5\%$ of $J_1$~\cite{Mertz2005, Lorenz2009, Zoghlin2023}. Therefore, we assume a small $J_1$ XXZ anisotropy with $\Delta_1=0.95$. Given that $J_2/\abs{J_1}=0.332 (>1/4)$, each $J_1$-$J_2$ chain would remain in the spiral phase if the interchain couplings were absent. However, the introduction of small interchain couplings $J_{\rm a}=1$ K and $J_{\rm b}=9$ K results in the system exhibiting a CAFO. It has been estimated that a $J_{\rm b}$ of $8.2$ K is required to induce CAFO~\cite{Li2CuO2_2011}, and the actual interchain coupling slightly exceeds this value. As shown in Fig.~\ref{fig:lattice}b, this system has a three-dimensional structure where each chain is coupled to four neighboring chains through $J_{\rm a}$ and $J_{\rm b}$ (see Fig.~\ref{fig:lattice}d). Therefore, we perform our DMRG calculations using a $12 \times 8$-chain cluster, applying open boundary conditions along the chain direction and periodic boundary conditions perpendicular to the chains, as depicted in Fig.~\ref{fig:lattice}f. Using the aforementioned parameters, we confirm a CAFO state with $2\abs{(\langle S^z_i \rangle)}=0.965$. This magnitude closely matches the experimentally estimated Cu ion moment of $0.96(4)\mu_{\rm B}$ at $T=1.5$ K~\cite{Sapina1990}, further supported by ab-initio calculations~\cite{Weht1998, Chung2003, Mertz2005}.

We then examine the results of the MES calculations using DDMRG method. Fig.~\ref{fig:MESmaterials}a plots $G_p(\omega)$ for $p$ values from 1 to 4. Focusing on a single $J_1$-$J_2$ chain, the FM order with a large magnetisation of $2\abs{\langle S^z_i \rangle}=0.965$ resembles a state almost fully saturated by an external magnetic field. Consequently, one might expect an MBS to form. The spectra clearly show that the $p=3$ spectrum $G_3(\omega)$ has the lowest excitation at $\omega \sim 2.58$ meV, indicating a triatic ground state. The energy gap between this peak and the lowest energy peak of $G_1(\omega)$ at $\omega \sim 2.73$ meV is $\Delta_{\rm gap,1} \sim 0.15$ meV, corresponding to the magnon binding energy. Note that our DDMRG calculations are performed at $T=0$. Additionally, $G_1(\omega)$ exhibits another gap, $\Delta_{\rm gap,2} \sim 1.41$ meV, between $\omega \sim 2.73$ meV and $\omega \sim 4.17$ meV, likely attributed to the spin gap induced by XXZ anisotropy. However, since the ground state is not a simple singlet, $\Delta_{\rm gap,2}$ is not a perfect gap, and some in-gap states are present between $\omega \sim 2.73$ meV and $\omega \sim 4.17$ meV.

For comparison, Fig.~\ref{fig:MESmaterials}b shows the experimental INS results near $k=0$. Given that the ground state ($\omega=0$) is a triatic state, the lowest energy peak of $G_3(\omega)$ corresponds to $\Delta E=0$. Meanwhile, $G_1(\omega)$ corresponds to the structure factor $S^{xx}(k,\omega)$ (or $S^{yy}(k,\omega)$), which is observable in INS experiments. Although the resolution around $\Delta E=0$ is limited in this INS data, a nearly-flat dispersion appears near $\Delta E=0.12$ meV (also see Fig. 2(c) in Ref.~\cite{Lorenz2009}), corresponding to our calculated $\Delta_{\rm gap,1}$ of $0.15$ meV. Furthermore, a gap of approximately $1.28$ meV exists from this nearly-flat dispersion to the lower bound of the FM dispersion, aligning with our calculated $\Delta_{\rm gap,2}$ and suggesting the presence of in-gap state weights. Note that this INS observation was made at $T=4.1$ K, while the critical temperature of the triatic state is estimated to be around $T_{\rm c}=1$ K by assuming the BCS prediction between energy gap at $T=0$ and the critical temperature, $\Delta(T=0)=1.76k_{\rm B}T_{\rm c}$~\cite{Tinkham2004}. Possibly, the gapped excitation, i.e., the lowest excitation peak $G_1(\omega)$ near $\Delta E=0.12$ meV, might be rather broadened. Therefore, more definitive evidence of MBS formation would require lower temperature and higher resolution experiments. Additionally, as demonstrated with the materials listed below, applying pressure to strengthen the CAFO could increase the MBS binding energy, potentially enhancing $\Delta_{\rm gap,1}$, observable through experimental measurements.

\subsection{Ca$_2$Y$_2$Cu$_5$O$_{10}$}\label{subsec:ca2y2cu5o10}

We next consider the compound Ca$_2$Y$_2$Cu$_5$O$_{10}$~\cite{Davies1991}. Its crystal structure is illustrated in Fig.~\ref{fig:lattice}c. This material exhibits a CAFO below $T_{\rm N}=29.5$ K, where spins align ferromagnetically along the a-axis (chain direction) and antiferromagnetically along the c-axis~\cite{Matsuda1999, Fong1999}. The ordered magnetic moment is $0.9 \mu_{\rm B}$. While the spins also align ferromagnetically along the b-axis, the interlayer couplings in the b direction are negligible compared to those within the ac layer. Thus, the ferromagnetic stacking of 2D CAFO planes can be regarded as the "2D analogue" of Li$_2$CuO$_2$.

However, a key difference from Li$_2$CuO$_2$ in the context of our MBS discussion is that the in-chain couplings in the $J_1$-$J_2$ chain for Ca$_2$Y$_2$Cu$_5$O$_{10}$ are $J_1 =-24$ meV and $J_2=5.5$ meV, leading to $J_2 /\abs{J_1}=0.23 (<0.25)$~\cite{Matsuda2019}. This implies that even in the absence of interchain couplings, each $J_1$-$J_2$ chain would order ferromagnetically. The CAFO in this material is driven by AFM interchain couplings, estimated as $J_{\rm a} + J_{\rm a} \sim 2.29$ meV~\cite{Kuzian2012, Matsuda2019}. Although the FM chains have their magnetisation directions fixed by the internal field, MBS states do not emerge.

Nevertheless, a notable point is that $J_2/\abs{J_1}=0.23$ is very close to the critical point of $0.25$. Therefore, tuning $J_2/\abs{J_1}$ to exceed $0.25$ may not be unrealistic. For instance, applying pressure/stretch might achieve this. Within the effective spin Hamiltonian, the smaller couplings are expected to be most sensitive to pressure. Consequently, it can be anticipated that the change in $J_2$ would be more significant than that in $J_1$ under pressure/stretch, making it feasible to achieve $J_2/\abs{J_1}>0.25$. Once $J_2/\abs{J_1}>0.25$ is realised, the CAFO could be retained by the relatively large interchain couplings $J_{\rm a} + J_{\rm a} \sim 2.29$ meV, potentially leading to the observation of MBS with a relatively large binding energy. Furthermore, if the value of $J_2/\abs{J_1}$ can be continuously varied by pressure, it might be possible to observe an MBS ground state that successively changes $p$ with pressure. Actually, a possible tuning of $J_2/\abs{J_1}$ by hydrostatic pressure has been suggested in Ref.~\cite{Caslin2016}.

\subsection{LiCuSbO$_4$}\label{subsec:licusbo4}

The third material under consideration is LiCuSbO$_4$~\cite{Dutton2012}. In this compound, the magnetic couplings between layers are negligible, allowing it to be effectively treated as a 2D system. The structure of the interchain couplings is equivalent to that depicted in Fig.~\ref{fig:lattice}e. The intrachain couplings have been estimated as $J_1=-125$ K and $J_2=37.6$ K~\cite{Grafe2017}, leading to a ratio of $J_2/\abs{J_1}=0.3$. Given this ratio is very close to the critical value of 0.25, even moderately strong AFM interchain couplings could stabilize a CAFO state. However, in practice, LiCuSbO$_4$ exhibits short-range incommensurate spin correlations below $T \sim 9$ K and, unlike the spiral spin-chain compounds LiCuVO$_4$~\cite{Gibson2004,Enderle2005} and LiCuZrO$_4$~\cite{Drechsler2007,Vavilova2009}, does not show long-range magnetic order down to $T \sim 0.1$ K. This may suggest that the interchain couplings in LiCuSbO$_4$ are very weak. Indeed, DFT and DFT+$U$ band structure calculations estimate $J_{\rm d} \sim 1$ K~\cite{Grafe2017}.

If the interchain couplings can be slightly enhanced through the application of pressure, CAFO might be realised, potentially leading to the manifestation of MBS. To explore this, we vary $J_{\rm d}$ around 1 K and calculate the MES $G_p(\omega)$ for a $24 \times 4$-chain cluster. Assuming an exchange anisotropy of approximately $10\%$~\cite{Dutton2012,Grafe2017,Bosiocic2017}, we set $\Delta_1=0.9$. The DDMRG results for $G_p(\omega)$ with $J_{\rm d}=0.75$ K and $1.25$ K, for $p$ values from 1 to 4, are shown in Fig.~\ref{fig:MESmaterials}c.

For $J_{\rm d}=0.75$ K, CAFO is not yet stabilised, and $G_1(\omega)$ exhibits the lowest energy excitation, indicating the absence of magnon binding. Nonetheless, the proximity of the lowest energy excitation in $G_3(\omega)$ suggests that a slight increase in $J_{\rm d}$ might stabilize a triatic ground state. Then, examining the results for $J_{\rm d}=1.25$ K, it is evident that $G_3(\omega)$ exhibits the lowest energy excitation, with a relatively large magnon binding energy of approximately 1 meV. Given that LiCuSbO$_4$ is already near the CAFO phase with $J_2/\abs{J_1}=0.3$, only a modest application of pressure might be required to observe both CAFO and MBS. Interestingly, as $J_{\rm d}$ increases, $G_1(\omega)$ shifts to higher energies while changing shape, whereas the spectra for $p>1$ remain largely unchanged. This observation supports the statement that the value of $p$ for the ground state generally corresponds to the number of magnon bindings stabilised by an external field in a single $J_1$-$J_2$ chain for a given $J_2/\abs{J_1}$.

\subsection{PbCuSO$_4$(OH)$_2$}\label{subsec:linarite}

The forth material we discuss is PbCuSO$_4$(OH)$_2$, commonly known as linarite~\cite{Effenberger1987}, which has a 2D layer lattice structure with interchain coupling $J_{\rm d}$, as illustrated in Fig.~\ref{fig:lattice}e. The in-chain couplings for this material have been estimated as $J_1=-100$ K and $J_2=36$ K~\cite{Willenberg2012}, with a ratio of $J_2/\abs{J_1}=0.36$, which is greater than the FM critical point of 0.25. Unlike LiCuSbO$_4$, linarite exhibits long-range spiral order with a small propagation number $q=0.186\pi$ along the $J_1$-$J_2$ chain below $T_{\rm N}=2.8$ K. This long-range order is believed to be stabilised by a sizable AFM interchain coupling, estimated to be $J_{\rm d}\sim 6$ K~\cite{Willenberg2016}. However, for this material to exhibit CAFO, an interchain coupling of approximately $J_{\rm d}=0.1\abs{J_1}\sim10$ K is required. Similar to LiCuSbO$_4$, there is potential for pressure effects to enhance the interchain coupling. If $J_{\rm d}>10$ K can be achieved, we can explore the possibility of realising an MBS ground state.

Given the highly anisotropic nature of PbCuSO$_4$(OH)$_2$, we set $\Delta_1=0.9$, though the actual anisotropy may be slightly larger. Fig.~\ref{fig:MESmaterials}d plots the MES $G_p(\omega)$ obtained from DDMRG calculations on a $24 \times 4$-chain cluster with $J_{\rm d}=15$ K. The lowest energy excitation state near $\omega=1.3$ meV is given by $G_2(\omega)$, indicating that the ground state is a nematic state. This observation aligns with the realisation of a nematic state in an external magnetic field for $J_2/\abs{J_1}=0.36$.

\subsection{Other materials}\label{subsec:others}

Other materials that potentially apply to our MBS emergence mechanism include CuSb$_2$O$_4$~\cite{Caslin2015} and Rb$_2$Cu$_2$Mo$_3$O$_{12}$~\cite{Hase2004}, which as also modeled by coupled FM-AFM $J_1$-$J_2$ chains. The ratios of their intrachain couplings are estimated to be $J_2/\abs{J_1} \sim 0.5$ and $J_2/\abs{J_1} = 0.37$, respectively. Due to the relatively small interchain couplings, these materials are thought to exhibit a spiral ordered state at low temperatures. As with the previously discussed materials, if these can transition to a CAFO state under the influence of pressure, MBS might be realised.

Another material worth mentioning is LiVCuO$_4$, frequently discussed as a candidate material for MBS under high external fields~\cite{Zhitomirsky2010,Svistov2011}. This material has extremely small interchain couplings, making it ideal for observing MBS at high magnetic fields~\cite{Stefan2011}. The ratio of intrachain couplings has not reached a complete consensus, with estimates ranging from $J_2/\abs{J_1} = 0.75$~\cite{Stefan2011, EPL2012}, to 1.42~\cite{Enderle2010}, and 3.49~\cite{Enderle2005}. In any case, these values are significantly distant from the FM critical point $J_2/\abs{J_1} = 0.25$, indicating that substantial interchain couplings would be necessary to induce a transition to CAFO. Therefore, a pressure-induced transition to CAFO might be unrealistic, and thus LiVCuO$_4$ can be eliminated from our list of candidate materials for our MBS mechanism.

Until now, we have considered systems where $J_1$-$J_2$ chains with a bipartite structure transition to CAFO under AFM interchain coupling. However, a simple bipartite structure is not that necessary. For instance, the Q1D mineral antlerite Cu$_3$SO$_4$(OH)$_4$. The magnetic model of this system can be represented as three-leg zigzag ladders, where the central leg hosts competing AFM exchanges, and each side leg features alternating NN FM exchanges $J_1$, $J_1^\prime$ and a weaker NNN AFM exchange $J_2$. These three chains are coupled by zigzag-shaped AFM interchain couplings. The side chains are dimerised FM-AFM $J_1$-$J_2$ chains, with parameters estimated as $J_1 = -26$, $J_1' = -11$, and $J_2 = -6$ (in meV) by GGA+U approximation~\cite{Anton2022}. Without interchain coupling, the side chains would be in a spiral state~\cite{Clio2017, Wardyn2023}. However, the experimentally observed low-temperature magnetic structure shows that the side legs are ferromagnetically polarised in opposite directions, while the central chain is in an AFM order. This suggests that the effect of interchain coupling causes the side chains to polarize ferromagnetically.

There is a possibility that MBS emerges due to the effect of interchain couplings in the side chains. Materials with such structures are diverse, indicating a wide potential for observing zero-field MBS, which has not been considered previously. We anticipate that our proposed mechanism for the emergence of MBSs will be of significant interest to experimentalists.

\section{Conclusion}\label{sec:conclusion}

In this study, we have demonstrated a novel mechanism for the stabilisation of MBS in Q1D cuprates through interchain interactions. Unlike conventional MBS that require an external magnetic field, our results show that AFM interchain coupling can induce MBS even in the absence of an external field. This finding is significant as it provides a new perspective on the interplay between interchain and intrachain interactions in low-dimensional magnetic systems.

Our numerical simulations using the DMRG method confirm that the AFM interchain coupling can act similarly to an internal magnetic field, leading to the formation of CAFO and the subsequent stabilisation of MBS. This is particularly intriguing because the AFM interchain coupling has a dual role; while it destabilizes MBS at high fields, it stabilizes them at zero field.

Notably, we found that MBS can be stabilised in Li$_2$CuO$_2$ under ambient pressure and zero magnetic field, highlighting the robustness of this mechanism. However, this aspect has not been experimentally investigated, as it has not been viewed from this perspective before. Additionally, we explored the conditions under which this phenomenon occurs in other materials such as Ca$_2$Y$_2$Cu$_5$O$_{10}$, LiCuSbO$_4$, and PbCuSO$_4$(OH)$_2$. Our findings suggest that the application of pressure or chemical substitution could further enhance the interchain coupling in these materials, making it feasible to experimentally observe the MBS predicted by our theoretical models.

Revisiting various physical phenomena from the perspective of effective internal magnetic fields due to interchain coupling may lead to new discoveries. This study serves as a promising example of such potential, demonstrating the significant implications for understanding magnetic excitations in Q1D systems.

\section*{Methods}\label{sec:methods}

\noindent
\textbf{Density-matrix renormalization group}

\noindent
We employ the DMRG technique~\cite{White1992} to study the ground-state properties. For single-chain calculations, we study periodic chains with lengths up to $l_x=192$ keeping $\chi=800$ density-matrix eigenstates for renormalization procedure. The discarded weight is negligible. When the magnon binding energy is calculated, it is better to use periodic chains to avoid picking up low excitations that may occur near the edges of open chains. For two-chain calculations, we study open ladders with lengths up to $l_x=280$ keeping $\chi=3600$. The largest discarded weight is $w_{\rm d} \sim 10^{-11}$. To explicitly fix the direction of magnetisation, which corresponds to the direction in which spin-rotation symmetry is broken in each chain, magnetic fields are applied to the sites at both ends of an open ladder. The direction of the magnetic fields is parallel to the z-axis, with fields at both ends of a single chain aligned in the same direction, and those on adjacent chains in opposite directions. However, the magnitude of magnetisation extrapolated to the thermodynamic limit is independent of the values of these edge fields. For $12 \times 8$-chain calculations, we keep $\chi=1200$ leading to $w_{\rm d} \sim 10^{-10}$. In this case, we focus on the collinear antiferromagnetic ordered state, where the quantum fluctuations are significantly suppressed, the DMRG accuracy is very good. Similar to the case of the open ladder, magnetic fields are applied to both ends of the system.\\

\noindent
\textbf{Dynamical density-matrix renormalization group} 

\noindent
We employ the dynamical DMRG (DDMRG) technique~\cite{eric2002} to calculate the magnon excitation spectrum (MES), defined as
We calculate the spin excitation spectrum
\begin{equation}
	G_p(k,\omega)=\frac{1}{\pi}{\rm Im}
	\langle 0|
	B^+_{-k} \frac{1}{\hat{H}+\omega-E_0-{\rm i}\eta}
	B^-_k
	|0 \rangle\, ,
	\label{spinspectrum}
\end{equation}
where $B^-_k=\sum_j \prod_{i=j}^{j+p-1} S^-_i \exp(ikr_j)$ for $p$-magnon flips, $|0\rangle$ and $E_0$ are, respectively, the wavefunction and energy of the ground state of the Hamiltonian \eqref{eq:ham}. A Lorentz distribution of width $\eta$ is used to broaden the delta peaks at the excited states. Since the DDMRG algorithm performs best for open boundary conditions, we study open clusters along the chain direction. When we calculate the local MES $G_p(\omega)=\sum_k G_p(k,\omega)$, the site position is fixed at $j=l_x/2$.

In two-chain calculations, we examine ladders of length $l_x=32$ with the broadening $\delta(\omega/p)=0.04\abs{J_1}$. The DDMRG approach is based on a variational principle, therefore it is necessary to prepare a 'good trial function' of the ground state with the density matrix eigenstates. We keep $\chi_1=1200$ in the first few dozen DMRG sweeps to obtain the ground state, and then to keep $\chi_2=400$ to calculate the excitation spectrum. This approach ensures that the maximum discarded weight is of the order of $10^{-6}$. For $24 \times 4$-chain calculations, the following parameters were set: $\chi_1=2000$, $\chi_2=600$, and $\eta/p=0.0025\abs{J_1}$, which resulted in a discarded weight of approximately $10^{-5}$. For $12 \times 8$-chain calculations, the following values were set: $\chi_1=1600$, $\chi_2=600$, and $\delta(\omega/p)=0.0025\abs{J_1}$. This resulted in a value of $w_{\rm d} \sim 10^{-7}$.

\backmatter

\bmhead{Supplementary information}

Finite-size effects in the MES are discussed in Supplemental Information.

\bmhead{Acknowledgments}

We thank Ulrike Nitzsche for technical assistance. We thank S.D.Wilson and R.O.Kuzian for fruitful discussion. This project is funded by the German Research Foundation (DFG) via the projects A05 of the Collaborative Research Center SFB
1143 (project-id 247310070). We  kindly  acknowledge  support  by  the  (Polish)  National  Science  Centre  (NCN, Poland)  under  Project  No.  2021/40/C/ST3/00177. 

\noindent
\textbf{Authors' contributions} 
C.E.A. and S.N. performed numerical calculations and analysed the results. S.N. designed the study and drafted the manuscript in discussion with C.E.A. and S.-L.D.. All authors approved the final manuscript.

\noindent
\textbf{Competing financial interests} The authors declare no competing interests.

\noindent
\bmhead{Data availability}
The data analyzed in the current study are available at https://doi.org/10.5281/zenodo.14499485





\bibliography{ICmagnonbound}

\begin{thebibliography}{10}
\expandafter\ifx\csname url\endcsname\relax
  \def\url#1{\burl{#1}}\fi
\expandafter\ifx\csname urlprefix\endcsname\relax\def\urlprefix{URL }\fi
\providecommand{\bibinfo}[2]{#2}
\providecommand{\eprint}[2][]{\url{#2}}
\providecommand{\doi}[1]{\url{https://doi.org/#1}}
\bibcommenthead

\bibitem{Bethe1931}
\bibinfo{author}{Bethe, H.}
\newblock \bibinfo{title}{Zur theorie der metalle}.
\newblock \emph{\bibinfo{journal}{Zeitschrift f{\"u}r Physik}}
  \textbf{\bibinfo{volume}{71}}, \bibinfo{pages}{205--226}
  (\bibinfo{year}{1931}).
\newblock \urlprefix\url{https://doi.org/10.1007/BF01341708}.

\bibitem{Wortis1963}
\bibinfo{author}{Wortis, M.}
\newblock \bibinfo{title}{Bound states of two spin waves in the heisenberg
  ferromagnet}.
\newblock \emph{\bibinfo{journal}{Phys. Rev.}} \textbf{\bibinfo{volume}{132}},
  \bibinfo{pages}{85--97} (\bibinfo{year}{1963}).
\newblock \urlprefix\url{https://link.aps.org/doi/10.1103/PhysRev.132.85}.

\bibitem{Hanus1963}
\bibinfo{author}{Hanus, J.}
\newblock \bibinfo{title}{Bound states in the heisenberg ferromagnet}.
\newblock \emph{\bibinfo{journal}{Phys. Rev. Lett.}}
  \textbf{\bibinfo{volume}{11}}, \bibinfo{pages}{336--338}
  (\bibinfo{year}{1963}).
\newblock \urlprefix\url{https://link.aps.org/doi/10.1103/PhysRevLett.11.336}.

\bibitem{Alicea2012}
\bibinfo{author}{Alicea, J.}
\newblock \bibinfo{title}{New directions in the pursuit of majorana fermions in
  solid state systems}.
\newblock \emph{\bibinfo{journal}{Reports on Progress in Physics}}
  \textbf{\bibinfo{volume}{75}}, \bibinfo{pages}{076501}
  (\bibinfo{year}{2012}).
\newblock \urlprefix\url{https://dx.doi.org/10.1088/0034-4885/75/7/076501}.

\bibitem{moore2010}
\bibinfo{author}{Moore, J.~E.}
\newblock \bibinfo{title}{The birth of topological insulators}.
\newblock \emph{\bibinfo{journal}{Nature}} \textbf{\bibinfo{volume}{464}},
  \bibinfo{pages}{194--198} (\bibinfo{year}{2010}).

\bibitem{Kruglyak2010}
\bibinfo{author}{Kruglyak, V.~V.}, \bibinfo{author}{Demokritov, S.~O.} \&
  \bibinfo{author}{Grundler, D.}
\newblock \bibinfo{title}{Magnonics}.
\newblock \emph{\bibinfo{journal}{Journal of Physics D: Applied Physics}}
  \textbf{\bibinfo{volume}{43}}, \bibinfo{pages}{264001}
  (\bibinfo{year}{2010}).
\newblock \urlprefix\url{https://dx.doi.org/10.1088/0022-3727/43/26/264001}.

\bibitem{Schaefer2020}
\bibinfo{author}{Sch{\"a}fer, F.}, \bibinfo{author}{Fukuhara, T.},
  \bibinfo{author}{Sugawa, S.}, \bibinfo{author}{Takasu, Y.} \&
  \bibinfo{author}{Takahashi, Y.}
\newblock \bibinfo{title}{Tools for quantum simulation with ultracold atoms in
  optical lattices}.
\newblock \emph{\bibinfo{journal}{Nature Reviews Physics}}
  \textbf{\bibinfo{volume}{2}}, \bibinfo{pages}{411--425}
  (\bibinfo{year}{2020}).
\newblock \urlprefix\url{https://doi.org/10.1038/s42254-020-0195-3}.

\bibitem{Fukuhara2013}
\bibinfo{author}{Fukuhara, T.} \emph{et~al.}
\newblock \bibinfo{title}{Microscopic observation of magnon bound states and
  their dynamics}.
\newblock \emph{\bibinfo{journal}{Nature}} \textbf{\bibinfo{volume}{502}},
  \bibinfo{pages}{76--79} (\bibinfo{year}{2013}).
\newblock \urlprefix\url{https://doi.org/10.1038/nature12541}.

\bibitem{Chubukov1991}
\bibinfo{author}{Chubukov, A.~V.}
\newblock \bibinfo{title}{Chiral, nematic, and dimer states in quantum spin
  chains}.
\newblock \emph{\bibinfo{journal}{Phys. Rev. B}} \textbf{\bibinfo{volume}{44}},
  \bibinfo{pages}{4693--4696} (\bibinfo{year}{1991}).
\newblock \urlprefix\url{https://link.aps.org/doi/10.1103/PhysRevB.44.4693}.

\bibitem{Heidrich-Meisner2006}
\bibinfo{author}{Heidrich-Meisner, F.}, \bibinfo{author}{Honecker, A.} \&
  \bibinfo{author}{Vekua, T.}
\newblock \bibinfo{title}{Frustrated ferromagnetic spin-$\frac{1}{2}$ chain in
  a magnetic field: The phase diagram and thermodynamic properties}.
\newblock \emph{\bibinfo{journal}{Phys. Rev. B}} \textbf{\bibinfo{volume}{74}},
  \bibinfo{pages}{020403} (\bibinfo{year}{2006}).
\newblock \urlprefix\url{https://link.aps.org/doi/10.1103/PhysRevB.74.020403}.

\bibitem{Vekua2007}
\bibinfo{author}{Vekua, T.}, \bibinfo{author}{Honecker, A.},
  \bibinfo{author}{Mikeska, H.-J.} \& \bibinfo{author}{Heidrich-Meisner, F.}
\newblock \bibinfo{title}{Correlation functions and excitation spectrum of the
  frustrated ferromagnetic spin-$\frac{1}{2}$ chain in an external magnetic
  field}.
\newblock \emph{\bibinfo{journal}{Phys. Rev. B}} \textbf{\bibinfo{volume}{76}},
  \bibinfo{pages}{174420} (\bibinfo{year}{2007}).
\newblock \urlprefix\url{https://link.aps.org/doi/10.1103/PhysRevB.76.174420}.

\bibitem{Kecke2007}
\bibinfo{author}{Kecke, L.}, \bibinfo{author}{Momoi, T.} \&
  \bibinfo{author}{Furusaki, A.}
\newblock \bibinfo{title}{Multimagnon bound states in the frustrated
  ferromagnetic one-dimensional chain}.
\newblock \emph{\bibinfo{journal}{Phys. Rev. B}} \textbf{\bibinfo{volume}{76}},
  \bibinfo{pages}{060407} (\bibinfo{year}{2007}).
\newblock \urlprefix\url{https://link.aps.org/doi/10.1103/PhysRevB.76.060407}.

\bibitem{Hikihara2008}
\bibinfo{author}{Hikihara, T.}, \bibinfo{author}{Kecke, L.},
  \bibinfo{author}{Momoi, T.} \& \bibinfo{author}{Furusaki, A.}
\newblock \bibinfo{title}{Vector chiral and multipolar orders in the
  spin-$\frac{1}{2}$ frustrated ferromagnetic chain in magnetic field}.
\newblock \emph{\bibinfo{journal}{Phys. Rev. B}} \textbf{\bibinfo{volume}{78}},
  \bibinfo{pages}{144404} (\bibinfo{year}{2008}).
\newblock \urlprefix\url{https://link.aps.org/doi/10.1103/PhysRevB.78.144404}.

\bibitem{Sudan2009}
\bibinfo{author}{Sudan, J.}, \bibinfo{author}{L\"uscher, A.} \&
  \bibinfo{author}{L\"auchli, A.~M.}
\newblock \bibinfo{title}{Emergent multipolar spin correlations in a
  fluctuating spiral: The frustrated ferromagnetic spin-$\frac{1}{2}$
  heisenberg chain in a magnetic field}.
\newblock \emph{\bibinfo{journal}{Phys. Rev. B}} \textbf{\bibinfo{volume}{80}},
  \bibinfo{pages}{140402} (\bibinfo{year}{2009}).
\newblock \urlprefix\url{https://link.aps.org/doi/10.1103/PhysRevB.80.140402}.

\bibitem{Zhitomirsky2010}
\bibinfo{author}{Zhitomirsky, M.~E.} \& \bibinfo{author}{Tsunetsugu, H.}
\newblock \bibinfo{title}{Magnon pairing in quantum spin nematic}.
\newblock \emph{\bibinfo{journal}{Europhysics Letters}}
  \textbf{\bibinfo{volume}{92}}, \bibinfo{pages}{37001} (\bibinfo{year}{2010}).
\newblock \urlprefix\url{https://dx.doi.org/10.1209/0295-5075/92/37001}.

\bibitem{Sato2013}
\bibinfo{author}{Sato, M.}, \bibinfo{author}{Hikihara, T.} \&
  \bibinfo{author}{Momoi, T.}
\newblock \bibinfo{title}{Spin-nematic and spin-density-wave orders in
  spatially anisotropic frustrated magnets in a magnetic field}.
\newblock \emph{\bibinfo{journal}{Phys. Rev. Lett.}}
  \textbf{\bibinfo{volume}{110}}, \bibinfo{pages}{077206}
  (\bibinfo{year}{2013}).
\newblock
  \urlprefix\url{https://link.aps.org/doi/10.1103/PhysRevLett.110.077206}.

\bibitem{Kuzian2023}
\bibinfo{author}{Kuzian, R.}
\newblock \bibinfo{title}{Methods of modeling of strongly correlated electron
  systems}.
\newblock \emph{\bibinfo{journal}{Nanomaterials}} \textbf{\bibinfo{volume}{13}}
  (\bibinfo{year}{2023}).
\newblock \urlprefix\url{https://www.mdpi.com/2079-4991/13/2/238}.

\bibitem{Svistov2011}
\bibinfo{author}{Svistov, L.} \emph{et~al.}
\newblock \bibinfo{title}{New high magnetic field phase of the frustrated s=
  1/2 chain compound licuvo 4}.
\newblock \emph{\bibinfo{journal}{JETP letters}} \textbf{\bibinfo{volume}{93}},
  \bibinfo{pages}{21--25} (\bibinfo{year}{2011}).

\bibitem{Mourigal2012}
\bibinfo{author}{Mourigal, M.} \emph{et~al.}
\newblock \bibinfo{title}{Evidence of a bond-nematic phase in
  ${\mathrm{licuvo}}_{4}$}.
\newblock \emph{\bibinfo{journal}{Phys. Rev. Lett.}}
  \textbf{\bibinfo{volume}{109}}, \bibinfo{pages}{027203}
  (\bibinfo{year}{2012}).
\newblock
  \urlprefix\url{https://link.aps.org/doi/10.1103/PhysRevLett.109.027203}.

\bibitem{Nawa2013}
\bibinfo{author}{Nawa, K.}, \bibinfo{author}{Takigawa, M.},
  \bibinfo{author}{Yoshida, M.} \& \bibinfo{author}{Yoshimura, K.}
\newblock \bibinfo{title}{Anisotropic spin fluctuations in the quasi
  one-dimensional frustrated magnet licuvo4}.
\newblock \emph{\bibinfo{journal}{Journal of the Physical Society of Japan}}
  \textbf{\bibinfo{volume}{82}}, \bibinfo{pages}{094709}
  (\bibinfo{year}{2013}).
\newblock \urlprefix\url{https://doi.org/10.7566/JPSJ.82.094709}.

\bibitem{Buettgen2014}
\bibinfo{author}{B\"uttgen, N.} \emph{et~al.}
\newblock \bibinfo{title}{Search for a spin-nematic phase in the
  quasi-one-dimensional frustrated magnet ${\mathrm{licuvo}}_{4}$}.
\newblock \emph{\bibinfo{journal}{Phys. Rev. B}} \textbf{\bibinfo{volume}{90}},
  \bibinfo{pages}{134401} (\bibinfo{year}{2014}).
\newblock \urlprefix\url{https://link.aps.org/doi/10.1103/PhysRevB.90.134401}.

\bibitem{Orlova2017}
\bibinfo{author}{Orlova, A.} \emph{et~al.}
\newblock \bibinfo{title}{Nuclear magnetic resonance signature of the
  spin-nematic phase in ${\mathrm{licuvo}}_{4}$ at high magnetic fields}.
\newblock \emph{\bibinfo{journal}{Phys. Rev. Lett.}}
  \textbf{\bibinfo{volume}{118}}, \bibinfo{pages}{247201}
  (\bibinfo{year}{2017}).
\newblock
  \urlprefix\url{https://link.aps.org/doi/10.1103/PhysRevLett.118.247201}.

\bibitem{Grafe2017}
\bibinfo{author}{Grafe, H.-J.} \emph{et~al.}
\newblock \bibinfo{title}{Signatures of a magnetic field-induced unconventional
  nematic liquid in the frustrated and anisotropic spin-chain cuprate
  licusbo4}.
\newblock \emph{\bibinfo{journal}{Scientific reports}}
  \textbf{\bibinfo{volume}{7}}, \bibinfo{pages}{6720} (\bibinfo{year}{2017}).

\bibitem{Ueda2009}
\bibinfo{author}{Ueda, H.~T.} \& \bibinfo{author}{Totsuka, K.}
\newblock \bibinfo{title}{Magnon bose-einstein condensation and various phases
  of three-dimensonal quantum helimagnets under high magnetic field}.
\newblock \emph{\bibinfo{journal}{Phys. Rev. B}} \textbf{\bibinfo{volume}{80}},
  \bibinfo{pages}{014417} (\bibinfo{year}{2009}).
\newblock \urlprefix\url{https://link.aps.org/doi/10.1103/PhysRevB.80.014417}.

\bibitem{PhysRevB.92.214415}
\bibinfo{author}{Nishimoto, S.}, \bibinfo{author}{Drechsler, S.-L.},
  \bibinfo{author}{Kuzian, R.}, \bibinfo{author}{Richter, J.} \&
  \bibinfo{author}{van~den Brink, J.}
\newblock \bibinfo{title}{Interplay of interchain interactions and exchange
  anisotropy: Stability and fragility of multipolar states in
  spin-$\frac{1}{2}$ quasi-one-dimensional frustrated helimagnets}.
\newblock \emph{\bibinfo{journal}{Phys. Rev. B}} \textbf{\bibinfo{volume}{92}},
  \bibinfo{pages}{214415} (\bibinfo{year}{2015}).
\newblock \urlprefix\url{https://link.aps.org/doi/10.1103/PhysRevB.92.214415}.

\bibitem{J_Phys_Conf_Ser_400_032069}
\bibinfo{author}{Nishimoto, S.}, \bibinfo{author}{Drechsler, S.-L.},
  \bibinfo{author}{Kuzian, R.~O.}, \bibinfo{author}{Richter, J.} \&
  \bibinfo{author}{van~den Brink, J.}
\newblock \bibinfo{title}{The effect of antiferromagnetic interchain coupling
  on multipolar phases in quasi-1d quantum helimagnets}.
\newblock \emph{\bibinfo{journal}{Journal of Physics: Conference Series}}
  \textbf{\bibinfo{volume}{400}}, \bibinfo{pages}{032069}
  (\bibinfo{year}{2012}).
\newblock \urlprefix\url{https://dx.doi.org/10.1088/1742-6596/400/3/032069}.

\bibitem{Chakravarty1991}
\bibinfo{author}{Chakravarty, S.}, \bibinfo{author}{Gelfand, M.~P.} \&
  \bibinfo{author}{Kivelson, S.}
\newblock \bibinfo{title}{Electronic correlation effects and superconductivity
  in doped fullerenes}.
\newblock \emph{\bibinfo{journal}{Science}} \textbf{\bibinfo{volume}{254}},
  \bibinfo{pages}{970--974} (\bibinfo{year}{1991}).
\newblock
  \urlprefix\url{https://www.science.org/doi/abs/10.1126/science.254.5034.970}.

\bibitem{Tonegawa1989}
\bibinfo{author}{Tonegawa, T.} \& \bibinfo{author}{Harada, I.}
\newblock \bibinfo{title}{One-dimensional isotropic spin-1/2 heisenberg magnet
  with ferromagnetic nearest-neighbor and antiferromagnetic
  next-nearest-neighbor interactions}.
\newblock \emph{\bibinfo{journal}{Journal of the Physical Society of Japan}}
  \textbf{\bibinfo{volume}{58}}, \bibinfo{pages}{2902--2915}
  (\bibinfo{year}{1989}).
\newblock \urlprefix\url{https://doi.org/10.1143/JPSJ.58.2902}.

\bibitem{Bursill1995}
\bibinfo{author}{Bursill, R.} \emph{et~al.}
\newblock \bibinfo{title}{Numerical and approximate analytical results for the
  frustrated spin- 1/2 quantum spin chain}.
\newblock \emph{\bibinfo{journal}{Journal of Physics: Condensed Matter}}
  \textbf{\bibinfo{volume}{7}}, \bibinfo{pages}{8605} (\bibinfo{year}{1995}).
\newblock \urlprefix\url{https://dx.doi.org/10.1088/0953-8984/7/45/016}.

\bibitem{Nersesyan1998}
\bibinfo{author}{Nersesyan, A.~A.}, \bibinfo{author}{Gogolin, A.~O.} \&
  \bibinfo{author}{E\ss{}ler, F. H.~L.}
\newblock \bibinfo{title}{Incommensurate spin correlations in spin- $1/2$
  frustrated two-leg heisenberg ladders}.
\newblock \emph{\bibinfo{journal}{Phys. Rev. Lett.}}
  \textbf{\bibinfo{volume}{81}}, \bibinfo{pages}{910--913}
  (\bibinfo{year}{1998}).
\newblock \urlprefix\url{https://link.aps.org/doi/10.1103/PhysRevLett.81.910}.

\bibitem{Itoi2001}
\bibinfo{author}{Itoi, C.} \& \bibinfo{author}{Qin, S.}
\newblock \bibinfo{title}{Strongly reduced gap in the zigzag spin chain with a
  ferromagnetic interchain coupling}.
\newblock \emph{\bibinfo{journal}{Phys. Rev. B}} \textbf{\bibinfo{volume}{63}},
  \bibinfo{pages}{224423} (\bibinfo{year}{2001}).
\newblock \urlprefix\url{https://link.aps.org/doi/10.1103/PhysRevB.63.224423}.

\bibitem{Sirker2011}
\bibinfo{author}{Sirker, J.} \emph{et~al.}
\newblock \bibinfo{title}{${J}_{1}\ensuremath{-}{J}_{2}$ heisenberg model at
  and close to its $z=4$ quantum critical point}.
\newblock \emph{\bibinfo{journal}{Phys. Rev. B}} \textbf{\bibinfo{volume}{84}},
  \bibinfo{pages}{144403} (\bibinfo{year}{2011}).
\newblock \urlprefix\url{https://link.aps.org/doi/10.1103/PhysRevB.84.144403}.

\bibitem{Furukawa2012}
\bibinfo{author}{Furukawa, S.}, \bibinfo{author}{Sato, M.},
  \bibinfo{author}{Onoda, S.} \& \bibinfo{author}{Furusaki, A.}
\newblock \bibinfo{title}{Ground-state phase diagram of a spin-$\frac{1}{2}$
  frustrated ferromagnetic xxz chain: Haldane dimer phase and gapped/gapless
  chiral phases}.
\newblock \emph{\bibinfo{journal}{Phys. Rev. B}} \textbf{\bibinfo{volume}{86}},
  \bibinfo{pages}{094417} (\bibinfo{year}{2012}).
\newblock \urlprefix\url{https://link.aps.org/doi/10.1103/PhysRevB.86.094417}.

\bibitem{Agrapidis2019}
\bibinfo{author}{Agrapidis, C.~E.}, \bibinfo{author}{Drechsler, S.-L.},
  \bibinfo{author}{van~den Brink, J.} \& \bibinfo{author}{Nishimoto, S.}
\newblock \bibinfo{title}{{Coexistence of valence-bond formation and
  topological order in the Frustrated Ferromagnetic $J_1$-$J_2$ Chain}}.
\newblock \emph{\bibinfo{journal}{SciPost Phys.}} \textbf{\bibinfo{volume}{6}},
  \bibinfo{pages}{019} (\bibinfo{year}{2019}).
\newblock \urlprefix\url{https://scipost.org/10.21468/SciPostPhys.6.2.019}.

\bibitem{Affleck1988}
\bibinfo{author}{Affleck, I.}, \bibinfo{author}{Kennedy, T.},
  \bibinfo{author}{Lieb, E.~H.} \& \bibinfo{author}{Tasaki, H.}
\newblock \bibinfo{title}{Valence bond ground states in isotropic quantum
  antiferromagnets}.
\newblock \emph{\bibinfo{journal}{Communications in Mathematical Physics}}
  \textbf{\bibinfo{volume}{115}}, \bibinfo{pages}{477--528}
  (\bibinfo{year}{1988}).

\bibitem{Haldane1983-1}
\bibinfo{author}{Haldane, F.}
\newblock \bibinfo{title}{Continuum dynamics of the 1-d heisenberg
  antiferromagnet: Identification with the o(3) nonlinear sigma model}.
\newblock \emph{\bibinfo{journal}{Physics Letters A}}
  \textbf{\bibinfo{volume}{93}}, \bibinfo{pages}{464--468}
  (\bibinfo{year}{1983}).
\newblock
  \urlprefix\url{https://www.sciencedirect.com/science/article/pii/037596018390631X}.

\bibitem{Haldane1983-2}
\bibinfo{author}{Haldane, F. D.~M.}
\newblock \bibinfo{title}{Nonlinear field theory of large-spin heisenberg
  antiferromagnets: Semiclassically quantized solitons of the one-dimensional
  easy-axis n\'eel state}.
\newblock \emph{\bibinfo{journal}{Phys. Rev. Lett.}}
  \textbf{\bibinfo{volume}{50}}, \bibinfo{pages}{1153--1156}
  (\bibinfo{year}{1983}).
\newblock \urlprefix\url{https://link.aps.org/doi/10.1103/PhysRevLett.50.1153}.

\bibitem{Oshikawa1992}
\bibinfo{author}{Oshikawa, M.}
\newblock \bibinfo{title}{Hidden z2*z2 symmetry in quantum spin chains with
  arbitrary integer spin}.
\newblock \emph{\bibinfo{journal}{Journal of Physics: Condensed Matter}}
  \textbf{\bibinfo{volume}{4}}, \bibinfo{pages}{7469} (\bibinfo{year}{1992}).
\newblock \urlprefix\url{https://dx.doi.org/10.1088/0953-8984/4/36/019}.

\bibitem{Sato2009}
\bibinfo{author}{Sato, M.}, \bibinfo{author}{Momoi, T.} \&
  \bibinfo{author}{Furusaki, A.}
\newblock \bibinfo{title}{Nmr relaxation rate and dynamical structure factors
  in nematic and multipolar liquids of frustrated spin chains under magnetic
  fields}.
\newblock \emph{\bibinfo{journal}{Phys. Rev. B}} \textbf{\bibinfo{volume}{79}},
  \bibinfo{pages}{060406} (\bibinfo{year}{2009}).
\newblock \urlprefix\url{https://link.aps.org/doi/10.1103/PhysRevB.79.060406}.

\bibitem{Momoi2024}
\bibinfo{author}{Momoi, T.}
\newblock \bibinfo{title}{Dynamics of quantum spin-nematics: Comparisons with
  canted antiferromagnets}.
\newblock \emph{\bibinfo{journal}{Phys. Rev. Res.}}
  \textbf{\bibinfo{volume}{6}}, \bibinfo{pages}{013169} (\bibinfo{year}{2024}).
\newblock
  \urlprefix\url{https://link.aps.org/doi/10.1103/PhysRevResearch.6.013169}.

\bibitem{Affleck1994}
\bibinfo{author}{Affleck, I.}, \bibinfo{author}{Gelfand, M.~P.} \&
  \bibinfo{author}{Singh, R. R.~P.}
\newblock \bibinfo{title}{A plane of weakly coupled heisenberg chains:
  theoretical arguments and numerical calculations}.
\newblock \emph{\bibinfo{journal}{Journal of Physics A: Mathematical and
  General}} \textbf{\bibinfo{volume}{27}}, \bibinfo{pages}{7313}
  (\bibinfo{year}{1994}).
\newblock \urlprefix\url{https://dx.doi.org/10.1088/0305-4470/27/22/009}.

\bibitem{Clio2019}
\bibinfo{author}{Agrapidis, C.~E.}, \bibinfo{author}{van~den Brink, J.} \&
  \bibinfo{author}{Nishimoto, S.}
\newblock \bibinfo{title}{Field-induced incommensurate ordering in heisenberg
  chains coupled by ising interaction: Model for ytterbium aluminum perovskite
  ${\mathrm{ybalo}}_{3}$}.
\newblock \emph{\bibinfo{journal}{Phys. Rev. B}} \textbf{\bibinfo{volume}{99}},
  \bibinfo{pages}{224423} (\bibinfo{year}{2019}).
\newblock \urlprefix\url{https://link.aps.org/doi/10.1103/PhysRevB.99.224423}.

\bibitem{Nikitin2021}
\bibinfo{author}{Nikitin, S.~E.} \emph{et~al.}
\newblock \bibinfo{title}{Multiple fermion scattering in the weakly coupled
  spin-chain compound ybalo3}.
\newblock \emph{\bibinfo{journal}{Nature communications}}
  \textbf{\bibinfo{volume}{12}}, \bibinfo{pages}{3599} (\bibinfo{year}{2021}).

\bibitem{EPL2012}
\bibinfo{author}{Nishimoto, S.} \emph{et~al.}
\newblock \bibinfo{title}{The strength of frustration and quantum fluctuations
  in livcuo4}.
\newblock \emph{\bibinfo{journal}{Europhysics Letters}}
  \textbf{\bibinfo{volume}{98}}, \bibinfo{pages}{37007} (\bibinfo{year}{2012}).

\bibitem{Lorenz2009}
\bibinfo{author}{Lorenz, W. E.~A.} \emph{et~al.}
\newblock \bibinfo{title}{Highly dispersive spin excitations in the chain
  cuprate li2cuo2}.
\newblock \emph{\bibinfo{journal}{Europhysics Letters}}
  \textbf{\bibinfo{volume}{88}}, \bibinfo{pages}{37002} (\bibinfo{year}{2009}).
\newblock \urlprefix\url{https://dx.doi.org/10.1209/0295-5075/88/37002}.

\bibitem{Ohta1993}
\bibinfo{author}{Ohta, H.} \emph{et~al.}
\newblock \bibinfo{title}{Epr and afmr of li2cuo2 in submillimeter wave
  region}.
\newblock \emph{\bibinfo{journal}{Journal of the Physical Society of Japan}}
  \textbf{\bibinfo{volume}{62}}, \bibinfo{pages}{785--792}
  (\bibinfo{year}{1993}).
\newblock \urlprefix\url{https://doi.org/10.1143/JPSJ.62.785}.

\bibitem{Mertz2005}
\bibinfo{author}{Mertz, D.}, \bibinfo{author}{Hayn, R.},
  \bibinfo{author}{Opahle, I.} \& \bibinfo{author}{Rosner, H.}
\newblock \bibinfo{title}{Calculated magnetocrystalline anisotropy and magnetic
  moment distribution in ${\mathrm{li}}_{2}{\mathrm{cuo}}_{2}$}.
\newblock \emph{\bibinfo{journal}{Phys. Rev. B}} \textbf{\bibinfo{volume}{72}},
  \bibinfo{pages}{085133} (\bibinfo{year}{2005}).
\newblock \urlprefix\url{https://link.aps.org/doi/10.1103/PhysRevB.72.085133}.

\bibitem{Zoghlin2023}
\bibinfo{author}{Zoghlin, E.}, \bibinfo{author}{Stone, M.~B.} \&
  \bibinfo{author}{Wilson, S.~D.}
\newblock \bibinfo{title}{Refined spin-wave model and multimagnon bound states
  in ${\mathrm{li}}_{2}{\mathrm{cuo}}_{2}$}.
\newblock \emph{\bibinfo{journal}{Phys. Rev. B}}
  \textbf{\bibinfo{volume}{108}}, \bibinfo{pages}{064408}
  (\bibinfo{year}{2023}).
\newblock \urlprefix\url{https://link.aps.org/doi/10.1103/PhysRevB.108.064408}.

\bibitem{Li2CuO2_2011}
\bibinfo{author}{Nishimoto, S.} \emph{et~al.}
\newblock \bibinfo{title}{Saturation field of frustrated chain cuprates: Broad
  regions of predominant interchain coupling}.
\newblock \emph{\bibinfo{journal}{Phys. Rev. Lett.}}
  \textbf{\bibinfo{volume}{107}}, \bibinfo{pages}{097201}
  (\bibinfo{year}{2011}).
\newblock
  \urlprefix\url{https://link.aps.org/doi/10.1103/PhysRevLett.107.097201}.

\bibitem{Sapina1990}
\bibinfo{author}{{n}a, F.~S.} \emph{et~al.}
\newblock \bibinfo{title}{Crystal and magnetic structure of li2cuo2}.
\newblock \emph{\bibinfo{journal}{Solid State Communications}}
  \textbf{\bibinfo{volume}{74}}, \bibinfo{pages}{779--784}
  (\bibinfo{year}{1990}).
\newblock
  \urlprefix\url{https://www.sciencedirect.com/science/article/pii/0038109890909344}.

\bibitem{Weht1998}
\bibinfo{author}{Weht, R.} \& \bibinfo{author}{Pickett, W.~E.}
\newblock \bibinfo{title}{Extended moment formation and second neighbor
  coupling in ${\mathrm{li}}_{2}{\mathrm{cuo}}_{2}$}.
\newblock \emph{\bibinfo{journal}{Phys. Rev. Lett.}}
  \textbf{\bibinfo{volume}{81}}, \bibinfo{pages}{2502--2505}
  (\bibinfo{year}{1998}).
\newblock \urlprefix\url{https://link.aps.org/doi/10.1103/PhysRevLett.81.2502}.

\bibitem{Chung2003}
\bibinfo{author}{Chung, E. M.~L.}, \bibinfo{author}{McIntyre, G.~J.},
  \bibinfo{author}{Paul, D.~M.}, \bibinfo{author}{Balakrishnan, G.} \&
  \bibinfo{author}{Lees, M.~R.}
\newblock \bibinfo{title}{Oxygen moment formation and canting in
  ${\mathrm{li}}_{2}{\mathrm{cuo}}_{2}$}.
\newblock \emph{\bibinfo{journal}{Phys. Rev. B}} \textbf{\bibinfo{volume}{68}},
  \bibinfo{pages}{144410} (\bibinfo{year}{2003}).
\newblock \urlprefix\url{https://link.aps.org/doi/10.1103/PhysRevB.68.144410}.

\bibitem{Tinkham2004}
\bibinfo{author}{Tinkham, M.}
\newblock \emph{\bibinfo{title}{Introduction to Superconductivity 2nd edn}}
  (\bibinfo{publisher}{McGraw-Hill, New York, 1975}, \bibinfo{year}{2004}).

\bibitem{Davies1991}
\bibinfo{author}{Davies, P.~K.}, \bibinfo{author}{Caignol, E.} \&
  \bibinfo{author}{King, T.}
\newblock \bibinfo{title}{New phases in the cao-m2o3-cuo (m = nd, gd, y)
  systems at 1000$^\circ$c}.
\newblock \emph{\bibinfo{journal}{Journal of the American Ceramic Society}}
  \textbf{\bibinfo{volume}{74}}, \bibinfo{pages}{569--573}
  (\bibinfo{year}{1991}).
\newblock
  \urlprefix\url{https://ceramics.onlinelibrary.wiley.com/doi/abs/10.1111/j.1151-2916.1991.tb04061.x}.

\bibitem{Matsuda1999}
\bibinfo{author}{Matsuda, M.}, \bibinfo{author}{Ohoyama, K.} \&
  \bibinfo{author}{Ohashi, M.}
\newblock \bibinfo{title}{Magnetic ordering of the edge-sharing cuo 2 chains in
  ca 2y 2cu 5o 10}.
\newblock \emph{\bibinfo{journal}{Journal of the Physical Society of Japan}}
  \textbf{\bibinfo{volume}{68}}, \bibinfo{pages}{269--272}
  (\bibinfo{year}{1999}).
\newblock \urlprefix\url{https://doi.org/10.1143/JPSJ.68.269}.

\bibitem{Fong1999}
\bibinfo{author}{Fong, H.~F.}, \bibinfo{author}{Keimer, B.},
  \bibinfo{author}{Lynn, J.~W.}, \bibinfo{author}{Hayashi, A.} \&
  \bibinfo{author}{Cava, R.~J.}
\newblock \bibinfo{title}{Spin structure of the dopable quasi-one-dimensional
  copper oxide
  ${\mathrm{ca}}_{2}{\mathrm{y}}_{2}{\mathrm{cu}}_{5}{\mathrm{o}}_{10}$}.
\newblock \emph{\bibinfo{journal}{Phys. Rev. B}} \textbf{\bibinfo{volume}{59}},
  \bibinfo{pages}{6873--6876} (\bibinfo{year}{1999}).
\newblock \urlprefix\url{https://link.aps.org/doi/10.1103/PhysRevB.59.6873}.

\bibitem{Matsuda2019}
\bibinfo{author}{Matsuda, M.} \emph{et~al.}
\newblock \bibinfo{title}{Highly dispersive magnons with spin-gap-like features
  in the frustrated ferromagnetic $s=\frac{1}{2}$ chain compound
  ${\mathrm{ca}}_{2}{\mathrm{y}}_{2}{\mathrm{cu}}_{5}{\mathrm{o}}_{10}$
  detected by inelastic neutron scattering}.
\newblock \emph{\bibinfo{journal}{Phys. Rev. B}}
  \textbf{\bibinfo{volume}{100}}, \bibinfo{pages}{104415}
  (\bibinfo{year}{2019}).
\newblock \urlprefix\url{https://link.aps.org/doi/10.1103/PhysRevB.100.104415}.

\bibitem{Kuzian2012}
\bibinfo{author}{Kuzian, R.~O.} \emph{et~al.}
\newblock
  \bibinfo{title}{${\mathrm{ca}}_{2}{\mathrm{y}}_{2}{\mathrm{cu}}_{5}{\mathbf{o}}_{10}$:
  The first frustrated quasi-1d ferromagnet close to criticality}.
\newblock \emph{\bibinfo{journal}{Phys. Rev. Lett.}}
  \textbf{\bibinfo{volume}{109}}, \bibinfo{pages}{117207}
  (\bibinfo{year}{2012}).
\newblock
  \urlprefix\url{https://link.aps.org/doi/10.1103/PhysRevLett.109.117207}.

\bibitem{Caslin2016}
\bibinfo{author}{Caslin, K.} \emph{et~al.}
\newblock \bibinfo{title}{Competing jahn-teller distortions and hydrostatic
  pressure effects in the quasi-one-dimensional quantum ferromagnet
  ${\mathrm{cuas}}_{2}{\mathrm{o}}_{4}$}.
\newblock \emph{\bibinfo{journal}{Phys. Rev. B}} \textbf{\bibinfo{volume}{93}},
  \bibinfo{pages}{022301} (\bibinfo{year}{2016}).
\newblock \urlprefix\url{https://link.aps.org/doi/10.1103/PhysRevB.93.022301}.

\bibitem{Dutton2012}
\bibinfo{author}{Dutton, S.~E.} \emph{et~al.}
\newblock \bibinfo{title}{Quantum spin liquid in frustrated one-dimensional
  ${\mathrm{licusbo}}_{4}$}.
\newblock \emph{\bibinfo{journal}{Phys. Rev. Lett.}}
  \textbf{\bibinfo{volume}{108}}, \bibinfo{pages}{187206}
  (\bibinfo{year}{2012}).
\newblock
  \urlprefix\url{https://link.aps.org/doi/10.1103/PhysRevLett.108.187206}.

\bibitem{Gibson2004}
\bibinfo{author}{Gibson, B.}, \bibinfo{author}{Kremer, R.},
  \bibinfo{author}{Prokofiev, A.}, \bibinfo{author}{Assmus, W.} \&
  \bibinfo{author}{McIntyre, G.}
\newblock \bibinfo{title}{Incommensurate antiferromagnetic order in the s=12
  quantum chain compound licuvo4}.
\newblock \emph{\bibinfo{journal}{Physica B: Condensed Matter}}
  \textbf{\bibinfo{volume}{350}}, \bibinfo{pages}{E253--E256}
  (\bibinfo{year}{2004}).
\newblock
  \urlprefix\url{https://www.sciencedirect.com/science/article/pii/S0921452604002455}.
\newblock \bibinfo{note}{Proceedings of the Third European Conference on
  Neutron Scattering}.

\bibitem{Enderle2005}
\bibinfo{author}{Enderle, M.} \emph{et~al.}
\newblock \bibinfo{title}{Quantum helimagnetism of the frustrated spin-1/2
  chain licuvo4}.
\newblock \emph{\bibinfo{journal}{Europhysics Letters}}
  \textbf{\bibinfo{volume}{70}}, \bibinfo{pages}{237} (\bibinfo{year}{2005}).
\newblock \urlprefix\url{https://dx.doi.org/10.1209/epl/i2004-10484-x}.

\bibitem{Drechsler2007}
\bibinfo{author}{Drechsler, S.-L.} \emph{et~al.}
\newblock \bibinfo{title}{Frustrated cuprate route from antiferromagnetic to
  ferromagnetic spin-$\frac{1}{2}$ heisenberg chains:
  ${\mathrm{li}}_{2}{\mathrm{zrcuo}}_{4}$ as a missing link near the quantum
  critical point}.
\newblock \emph{\bibinfo{journal}{Phys. Rev. Lett.}}
  \textbf{\bibinfo{volume}{98}}, \bibinfo{pages}{077202}
  (\bibinfo{year}{2007}).
\newblock
  \urlprefix\url{https://link.aps.org/doi/10.1103/PhysRevLett.98.077202}.

\bibitem{Vavilova2009}
\bibinfo{author}{Vavilova, E.} \emph{et~al.}
\newblock \bibinfo{title}{Quantum electric dipole glass and frustrated
  magnetism near a critical point in li2zrcuo4}.
\newblock \emph{\bibinfo{journal}{Europhysics Letters}}
  \textbf{\bibinfo{volume}{88}}, \bibinfo{pages}{27001} (\bibinfo{year}{2009}).
\newblock \urlprefix\url{https://dx.doi.org/10.1209/0295-5075/88/27001}.

\bibitem{Bosiocic2017}
\bibinfo{author}{Bosio\ifmmode \check{c}\else
  \v{c}\fi{}i\ifmmode~\acute{c}\else \'{c}\fi{}, M.} \emph{et~al.}
\newblock \bibinfo{title}{Possible quadrupolar nematic phase in the frustrated
  spin chain ${\mathrm{licusbo}}_{4}$: An nmr investigation}.
\newblock \emph{\bibinfo{journal}{Phys. Rev. B}} \textbf{\bibinfo{volume}{96}},
  \bibinfo{pages}{224424} (\bibinfo{year}{2017}).
\newblock \urlprefix\url{https://link.aps.org/doi/10.1103/PhysRevB.96.224424}.

\bibitem{Effenberger1987}
\bibinfo{author}{Effenberger, H.}
\newblock \bibinfo{title}{Crystal structure and chemical formula of
  schmiederite, pb2cu2 (oh) 4 (seo3)(seo4), with a comparison to linarite, pbcu
  (oh) 2 (so4)}.
\newblock \emph{\bibinfo{journal}{Mineralogy and petrology}}
  \textbf{\bibinfo{volume}{36}}, \bibinfo{pages}{3--12} (\bibinfo{year}{1987}).

\bibitem{Willenberg2012}
\bibinfo{author}{Willenberg, B.} \emph{et~al.}
\newblock \bibinfo{title}{Magnetic frustration in a quantum spin chain: The
  case of linarite ${\mathrm{pbcuso}}_{4}(\mathrm{OH}{)}_{2}$}.
\newblock \emph{\bibinfo{journal}{Phys. Rev. Lett.}}
  \textbf{\bibinfo{volume}{108}}, \bibinfo{pages}{117202}
  (\bibinfo{year}{2012}).
\newblock
  \urlprefix\url{https://link.aps.org/doi/10.1103/PhysRevLett.108.117202}.

\bibitem{Willenberg2016}
\bibinfo{author}{Willenberg, B.} \emph{et~al.}
\newblock \bibinfo{title}{Complex field-induced states in linarite
  ${\mathrm{pbcuso}}_{4}(\mathrm{OH}{)}_{2}$ with a variety of high-order
  exotic spin-density wave states}.
\newblock \emph{\bibinfo{journal}{Phys. Rev. Lett.}}
  \textbf{\bibinfo{volume}{116}}, \bibinfo{pages}{047202}
  (\bibinfo{year}{2016}).
\newblock
  \urlprefix\url{https://link.aps.org/doi/10.1103/PhysRevLett.116.047202}.

\bibitem{Caslin2015}
\bibinfo{author}{Caslin, K.}
\newblock \bibinfo{title}{Investigation of frustrated quasi-one-dimensional
  quantum spin-chain materials}.
\newblock \emph{\bibinfo{journal}{Ph.D. Thesis}}  (\bibinfo{year}{2015}).

\bibitem{Hase2004}
\bibinfo{author}{Hase, M.} \emph{et~al.}
\newblock \bibinfo{title}{Magnetic properties of
  ${\mathrm{rb}}_{2}{\mathrm{cu}}_{2}{\mathrm{mo}}_{3}{\mathrm{o}}_{12}$
  including a one-dimensional spin-$1/2$ heisenberg system with ferromagnetic
  first-nearest-neighbor and antiferromagnetic second-nearest-neighbor exchange
  interactions}.
\newblock \emph{\bibinfo{journal}{Phys. Rev. B}} \textbf{\bibinfo{volume}{70}},
  \bibinfo{pages}{104426} (\bibinfo{year}{2004}).
\newblock \urlprefix\url{https://link.aps.org/doi/10.1103/PhysRevB.70.104426}.

\bibitem{Stefan2011}
\bibinfo{author}{Drechsler, S.-L.} \emph{et~al.}
\newblock \bibinfo{title}{Comment on ``two-spinon and four-spinon continuum in
  a frustrated ferromagnetic spin-$1/2$ chain''}.
\newblock \emph{\bibinfo{journal}{Phys. Rev. Lett.}}
  \textbf{\bibinfo{volume}{106}}, \bibinfo{pages}{219701}
  (\bibinfo{year}{2011}).
\newblock
  \urlprefix\url{https://link.aps.org/doi/10.1103/PhysRevLett.106.219701}.

\bibitem{Enderle2010}
\bibinfo{author}{Enderle, M.} \emph{et~al.}
\newblock \bibinfo{title}{Two-spinon and four-spinon continuum in a frustrated
  ferromagnetic spin-$1/2$ chain}.
\newblock \emph{\bibinfo{journal}{Phys. Rev. Lett.}}
  \textbf{\bibinfo{volume}{104}}, \bibinfo{pages}{237207}
  (\bibinfo{year}{2010}).
\newblock
  \urlprefix\url{https://link.aps.org/doi/10.1103/PhysRevLett.104.237207}.

\bibitem{Anton2022}
\bibinfo{author}{Kulbakov, A.~A.} \emph{et~al.}
\newblock \bibinfo{title}{Coupled frustrated ferromagnetic and
  antiferromagnetic quantum spin chains in the quasi-one-dimensional mineral
  antlerite ${\mathrm{cu}}_{3}{\mathrm{so}}_{4}$(oh)${}_{4}$}.
\newblock \emph{\bibinfo{journal}{Phys. Rev. B}}
  \textbf{\bibinfo{volume}{106}}, \bibinfo{pages}{L020405}
  (\bibinfo{year}{2022}).
\newblock
  \urlprefix\url{https://link.aps.org/doi/10.1103/PhysRevB.106.L020405}.

\bibitem{Clio2017}
\bibinfo{author}{Agrapidis, C.~E.}, \bibinfo{author}{Drechsler, S.-L.},
  \bibinfo{author}{van~den Brink, J.} \& \bibinfo{author}{Nishimoto, S.}
\newblock \bibinfo{title}{Crossover from an incommensurate singlet spiral state
  with a vanishingly small spin gap to a valence-bond solid state in dimerized
  frustrated ferromagnetic spin chains}.
\newblock \emph{\bibinfo{journal}{Phys. Rev. B}} \textbf{\bibinfo{volume}{95}},
  \bibinfo{pages}{220404} (\bibinfo{year}{2017}).
\newblock \urlprefix\url{https://link.aps.org/doi/10.1103/PhysRevB.95.220404}.

\bibitem{Wardyn2023}
\bibinfo{author}{Wardyn, J. m.~k.}, \bibinfo{author}{Nishimoto, S.} \&
  \bibinfo{author}{Agrapidis, C.~E.}
\newblock \bibinfo{title}{Existence of two distinct valence bond solid states
  in the dimerized frustrated ferromagnetic
  ${J}_{1}\text{\ensuremath{-}}{J}_{1}^{\ensuremath{'}}\text{\ensuremath{-}}{J}_{2}$
  chain}.
\newblock \emph{\bibinfo{journal}{Phys. Rev. B}}
  \textbf{\bibinfo{volume}{108}}, \bibinfo{pages}{205111}
  (\bibinfo{year}{2023}).
\newblock \urlprefix\url{https://link.aps.org/doi/10.1103/PhysRevB.108.205111}.

\bibitem{White1992}
\bibinfo{author}{White, S.~R.}
\newblock \bibinfo{title}{Density matrix formulation for quantum
  renormalization groups}.
\newblock \emph{\bibinfo{journal}{Phys. Rev. Lett.}}
  \textbf{\bibinfo{volume}{69}}, \bibinfo{pages}{2863--2866}
  (\bibinfo{year}{1992}).
\newblock \urlprefix\url{https://link.aps.org/doi/10.1103/PhysRevLett.69.2863}.

\bibitem{eric2002}
\bibinfo{author}{Jeckelmann, E.}
\newblock \bibinfo{title}{Dynamical density-matrix renormalization-group
  method}.
\newblock \emph{\bibinfo{journal}{Phys. Rev. B}} \textbf{\bibinfo{volume}{66}},
  \bibinfo{pages}{045114} (\bibinfo{year}{2002}).
\newblock \urlprefix\url{https://link.aps.org/doi/10.1103/PhysRevB.66.045114}.

\end{thebibliography}


\end{document}


\widetext

\large
\begin{center}
{\bf SUPPLEMENTARY INFORMATION for\\
Self-organised magnon condensation in quasi-1D edge-shared cuprates without external fields}

\normalsize
\vspace{3.0mm}
Cli\`o Efthimia Agrapidis$^1$, Stefan-Ludwig Drechsler$^2$, and Satoshi Nishimoto$^{2,3}$

\small

\vspace{1.5mm}
{\it
$^1$Institute of Theoretical Physics, Faculty of Physics,
University of Warsaw, Pasteura 5, PL-02093 Warsaw, Poland\\
$^2$IFW Dresden, Helmholtzstra\ss e 20, 01069 Dresden, Germany\\
$^3$Department of Physics, TU Dresden, 01069 Dresden, Germany\\
}

\end{center}

\normalsize

\renewcommand{\theequation}{S\arabic{equation}}
\renewcommand{\thefigure}{S\arabic{figure}}
\renewcommand{\thetable}{S\arabic{table}}
\setcounter{equation}{0}
\setcounter{figure}{0}
\setcounter{table}{0}

\date{\today}

\subsection{Finite-size effects in the magnon excitation spectrum}

\begin{figure}[ht]
	\centering
	\includegraphics[width=0.65\linewidth]{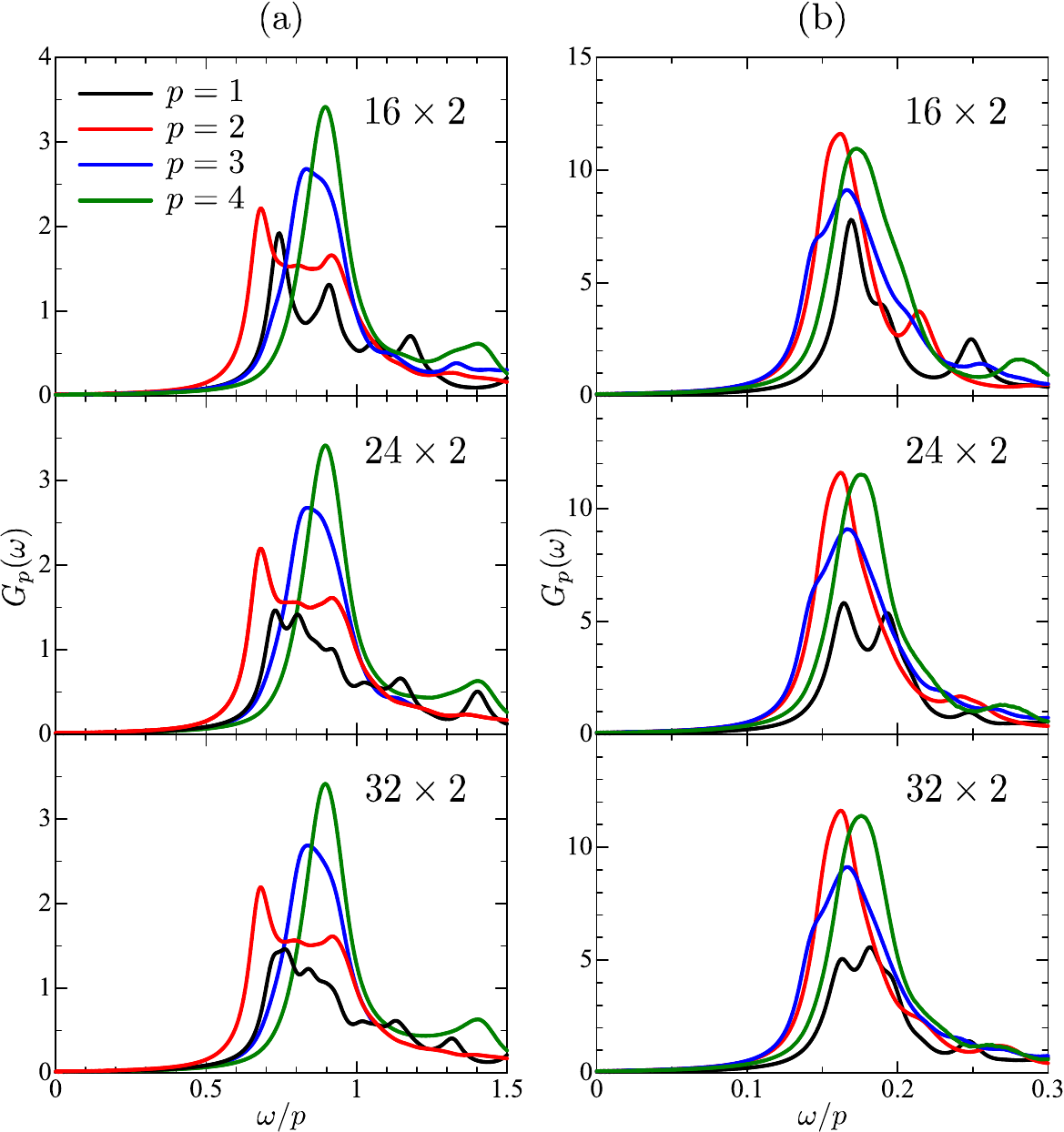}
	\caption{
		DDMRG results for the MES $G_p(\omega)$ at various values of $p=1$, $2$, $3$, and $4$, for (a) $J_2/\abs{J_1}=1/2$. $\Delta_{\rm a}=0.1$, $J_{\rm a}/\abs{J_1}=1.0$ and (b) $J_2/\abs{J_1}=1/3$, $\Delta_{\rm a}=0.1$, $J_{\rm a}/\abs{J_1}=0.2$, with the XXZ anisotropy applied to interchain couplings $J_{\rm a}$. Results with different cluster sizes  
		$16 \times 2$, $24 \times 2$, and $32 \times 2$ are compared.
	}
	\label{fig_size_dep}
\end{figure}

In Fig.~4b,c of the main text, we present the DDMRG results for the magnon excitation spectrum (MES) using a $32 \times 2$ open cluster. It is well established that open clusters are more feasible for calculating spectral functions because finite-size effects have a much smaller influence on the results compared to using periodic clusters. Here, we verify whether the finite-size effects are indeed within an acceptable range in our DDMRG calculations. In Fig.~\ref{fig_size_dep}, we present a comparison of the magnon excitation spectrum (MES) results obtained using different cluster sizes, ranging from $16 \times 2$ to $32 \times 2$. The parameters used are identical to those employed in Fig.~4c of the main text. While the spectral shape exhibits slight variations with cluster size, the behaviour near the onset of the spectrum is largely independent of cluster size. Consequently, it is feasible to identify low-energy excitations corresponding to the magnon bound state with high resolution even in relatively small clusters.

